\documentclass[trackchanges]{aastex701}

\usepackage{multirow}%
\usepackage{amsmath,amssymb,amsfonts}%
\usepackage{bm}
\usepackage{amsthm}%
\usepackage{mathrsfs}%
\usepackage[title]{appendix}%
\usepackage{xcolor}%
\usepackage{textcomp}%
\usepackage{algorithm}%
\usepackage{algorithmicx}%
\usepackage{algpseudocode}%
\usepackage{listings}%
\usepackage{filecontents}
\usepackage{multibib}
\usepackage[left]{lineno}
\usepackage{soul}
\usepackage{comment}
\renewcommand{\hl}[1]{#1} %silent all highlights

\newcites{supp}{Supplementary References}

\usepackage{graphicx}
\usepackage{subcaption}
\usepackage{threeparttable}
\usepackage{tabularx}
\usepackage{float}

\usepackage{natbib}

\begin{document}

\title{Soot Planets instead of Water Worlds}

%\author{Jie Li, Edwin A. Bergin, Marc M. Hirschmann, Geoffrey A. Blake, Fred J. Ciesla, Eliza M.-R. Kempton}

\author[orcid=0000-0000-0000-0001,sname='North America']{Jie Li}
\affiliation{Department of Earth and Environmental Sciences, University of Michigan, Ann Arbor, MI 48109, USA}
\email[show]{jackieli@umich.edu, ebergin@umich.edu}  

\author[orcid=0000-0000-0000-0001,sname='North America']{Edwin A. Bergin}
\affiliation{Department of Astronomy, University of Michigan, Ann Arbor, MI 48109, USA}
\email{ebergin@umich.edu}

\author[orcid=0000-0000-0000-0001,sname='North America']{Marc M. Hirschmann}
\affiliation{Department of Earth and Environmental Sciences, University of Minnesota, Minneapolis, MN 55455, USA}
\email{ebergin@umich.edu}

\author[orcid=0000-0000-0000-0001,sname='North America']{Geoffrey A. Blake}
\affiliation{Division of Geological \& Planetary Sciences, California Institute of Technology, Pasadena, CA 91125, USA}
\email{ebergin@umich.edu}

\author[orcid=0000-0000-0000-0001,sname='North America']{Fred J. Ciesla}
\affiliation{Department of Geophysical Sciences, University of Chicago, Chicago, IL 60637, USA} 
\email{ebergin@umich.edu}

\author[orcid=0000-0000-0000-0001,sname='North America']{Eliza M.-R. Kempton}
\affiliation{Department of Astronomy and Astrophysics, University of Chicago, Chicago, IL 60637, USA} 
\email{ebergin@umich.edu}

\begin{abstract}
Some low-density exoplanets are thought to be water-rich worlds that formed beyond the snow line of their protoplanetary disc, possibly accreting coequal portions of rock and water \citep{Luque22, Bitsch2019_waterworld, Izidoro2021}. However, the compositions of bodies within the Solar System and the stability of volatile-rich solids in accretionary disks suggest that a planet rich in water should also acquire as much as 40\% refractory organic carbon (``soot'') \citep{Li21, Bergin2023}.  This would reduce the water mass fraction well below 50\%, making the composition of these planets similar to those of Solar System comets \citep{Rubin19}. Here we show that soot-rich planets, with or without water, can account for the low average densities of exoplanets that were previously attributed to a binary combination of rock and water. Formed in locations beyond the soot and/or snow lines in disks, these planets are likely common in our galaxy and already observed by JWST. The surfaces and interiors of soot-rich planets will be influenced by the chemical and physical properties of carbonaceous phases, and the atmospheres of such planets may contain plentiful methane and other hydrocarbons, with implications for photochemical haze generation and habitability. 
\end{abstract}
 
%\keywords{Soot line, snow line, refractory carbon}

%\thispagestyle{empty}

%\section{Introduction}
%The demographics of exoplanetary systems show that the most common planet in the Milky Way has a size that is slightly larger than the Earth, but smaller than Neptune \citep[1--4~R$_\oplus$;][]{Borucki2011}. 

\section{Introduction}
The masses and radii of the most common planets in the observable galaxy are intermediate between Earth and Neptune \citep{Borucki2011, fulton17}.  These ``sub-Neptune'' exoplanets are typically less dense than the Earth \citep[e.g.,][]{Charbonneau2009, Luque22}, indicating that they contain abundant lighter components.  Although such planets can be explained by rocky cores surrounded by large H$_2$--He envelopes \citep{owen17, rogers2023conclusive}, they have alternatively been considered as ``water worlds'', with as much as 50\% H$_2$O by mass \citep[e.g.,][]{rogers10, Luque22}. Formation of water worlds are thought to be a natural consequence of planetary accretion outside the ``ice line'' or ``snow line'', where  water ice is abundant \citep{Mobidelli2000,Bitsch2019_waterworld}, and therefore bodies with both rock and water ice should readily form.  The possible existence of such sub-Neptune planets, which have been proposed to form a distinct population \citep{Luque22}, has elicited considerable inquiry, including the potential of temperate water worlds hosting liquid oceans \citep{kite2018habitability, madhu21}. 

\begin{figure}[h]
\centering
\includegraphics[width=\linewidth]{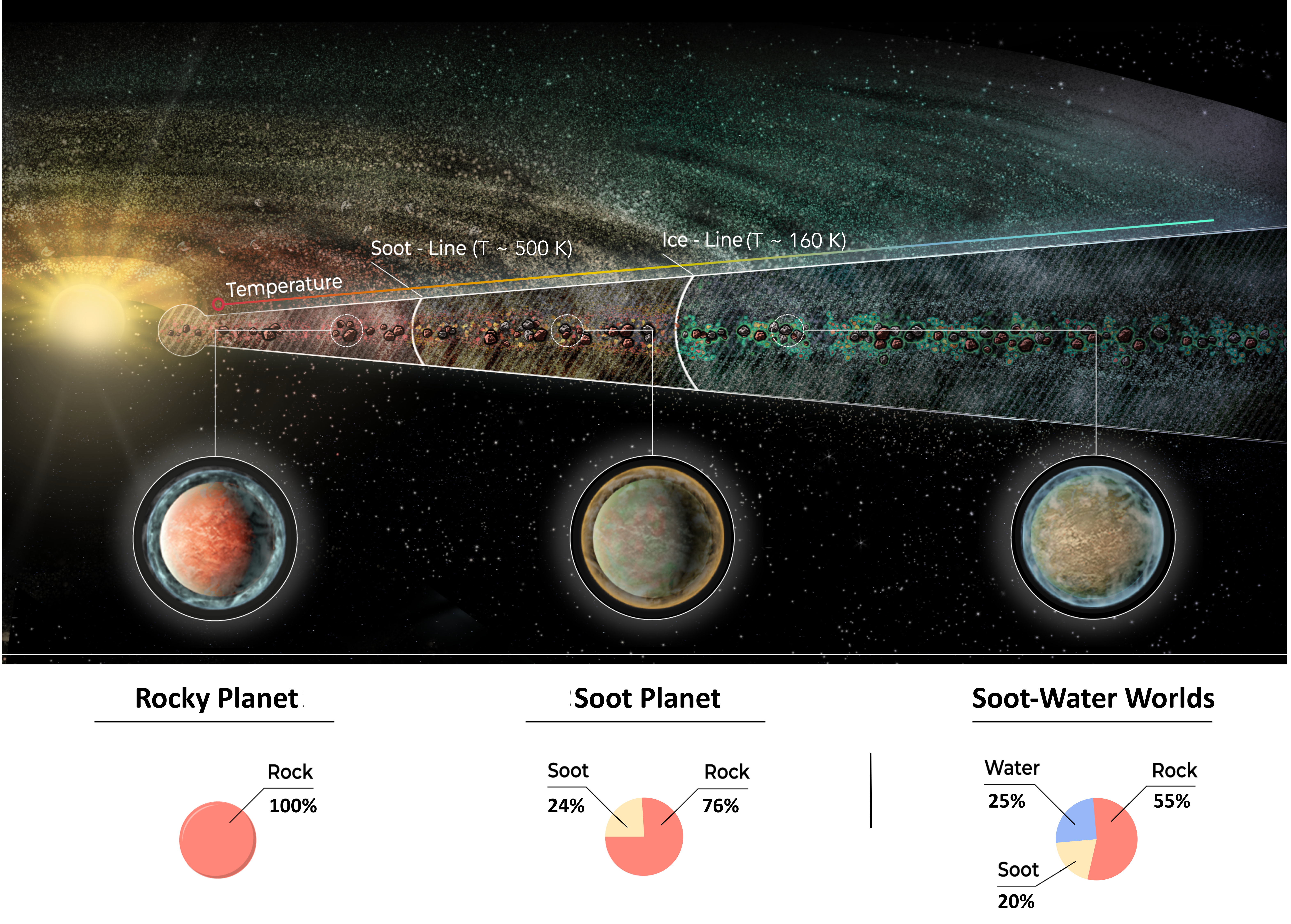}\caption{Illustration of the soot line and water ice line structure of protoplanetary disks, and the three chemically distinct planet archetypes. The water ice line is also known as the water snow line or snow line. Compositions of the model rocky planet, soot planet, and soot-water worlds are shown as pie charts, with red, yellow, and blue segments representing the mass fractions of rock, soot, and water, respectively (Table~\ref{Tab:composition}). For the soot-water worlds, the pie chart and numbers refer to the end-member dry case and the numbers for the end-member wet case are: 36\% rock, 14\% soot, and 50\% water. The soot line is defined by its sublimation temperature of 500 K under disk conditions \citep{Li21}. The water ice line is placed at 160 K \citep{Minissale22}. The locations of the soot lines and ice lines vary with time and across stellar mass range.  The soot line and ice line are drawn for illustrative purposes, Their actual shapes depend on the pressure distribution in the disk, as well as the intensity of viscous heating and the contribution from irradiation from the central star, which vary over time and with physical conditions in the disk. For this study, it is the relative locations that matter. Figure credit Ari Gea and Sayo Studio.}
\label{fig:cartoon}
\end{figure}

An alternative explanation for low-density sub-Neptunes, however, is that they are rich in carbonaceous materials, with or without appreciable water fractions. The conventional condensation sequence would not predict any appreciable carbon inside of the CO snow line, at about 30 AU \citep{Lodders03}. The origin of the carbon phases found in primitive meteorites (and Ryugu, Bennu, etc.) is debated, but it is clear that this relatively refractory material was abundant in small primitive objects in the early solar system and was available to accrete into planetesimals and therefore planets. Carbon-rich grains may be inherited from the interstellar medium or forged in protoplanetary disk atmospheres.  As posited in \citet{Li21}, the distribution and state of carbon in protoplanetary disks inside of the CO snow line is governed by irreversible sublimation instead of condensation. The ``soot line'' or ``tar line'' refers to the location where carbon-rich grains are destroyed via thermally driven reactions \citep{Kress2010,Lodders2004,Li21}. It follows that planets accreted in the inner disk and beyond the soot line would acquire a large amount of carbon in the form of refractory organics (Fig.~\ref{fig:cartoon}).

Indeed, protoplanetary disks are known to have a significant reservoir of refractory carbonaceous grains, or ``soot'', which refers to a chemical component consisting of many phases, including meteoritic insoluble organic matter (IOM) and carbonaceous materials found in comets \citep{Kress2010}. It is a term that describes solids rich in carbon (C), hydrogen (H), oxygen (O), and with appreciable nitrogen (N), collectively known as CHON. The specific carriers of the carbon may include a variety of species of refractory organics such as refractory polyaromatic organics and semi-volatile organics found at the surface of comet 67P/CG \citep{2016Icar..272...32Q}, refractory organics in the carbon-rich dust grains of comet 67P/CG and 1P/Halley \citep[][and references therein]{Bergin15, Rubin19}, and organic carbon carriers in ordinary chondrites \citep{Grady89a} and martian meteorites \citep{Sephton03}. Soot is stable and remains in the solid state to much greater temperatures ($\sim$500 K) \citep{Bergin15, Gail17,Li21} than water ice ($\sim$150 K) \citep{Lodders03}.  A significant soot component would therefore be present in the formation zone of planets accreting outside of the snow line, unless there were significant periods of high temperature events driven by time variable accretion.  The compositions of comets indicate that soot may make up to 40\% of solid mass in this region \citep{Bardyn2017, Rubin19}, while Uranus and Neptune are believed to be dominated by hydrocarbons rather than H$_2$O \citep{Malamud2024}. Finally, the high C/O ratio of Jupiter observed by Galileo may reflect the accretion of abundant refractory carbon \citep{Lodders2004}. 

Importantly, because refractory carbon survives to higher temperatures than water ice, there are also likely regions where planets accrete both rock and soot, but little water (Fig.~\ref{fig:cartoon}).  Inside of the soot line, only rocky solids would be present.  As a result, accretion of smaller gas-poor planets can potentially produce three generic archetypes depending on where accretion occurs: Rock-rich worlds with low carbon+water content (e.g., terrestrial planets), carbon-rich rocky worlds with low water content \citep{Bergin2023}, and rock/carbon/water worlds. In this study, we test the hypothesis of soot-rich planets against measured mass-radius relations. Furthermore, we discuss atmospheric characteristics for future detection of soot-rich planets and explore the implications of soot for planetary habitability.

%\newpage
\section{Planet Models}
We constructed models of exoplanet composition based on proto-solar elemental abundances and the distribution of solid materials found in comets (Fig.~\ref{fig:cartoon}), as described in detail in Appendix. With the primary building blocks formed inside of the soot line, ``rocky planets'' are assumed to be composed entirely of metal and silicate, analogous to Earth. ``Soot planets'', formed between the soot line and snow line, are made of both rock and soot. Planets forming outside of the water snow line are termed ``soot-water worlds" and  contain rock, soot, and water. We note that `soot-water worlds' include a significant component of hydrocarbon-rich material and are therefore distinct from the `water worlds' posited in earlier work \citep{Luque22,Bitsch2019_waterworld,rogers2023conclusive}. 

Computing the exact mass-radius relationships of super-Earths or sub-Neptunes requires detailed and robust knowledge of the relevant phases at the high pressures-temperature conditions of planetary interiors, which is still largely lacking. In particular, predicting the fate of soot in planets has significant uncertainties because the prevailing carbonaceous phases for a given set of planetary parameters are poorly constrained. To overcome this challenge, we choose to treat the soot component as a fictive phase contributing to the properties of the planets, as described in Appendix. Using studies of meteorites \citep{Alexander12,Bergin2023}, the "soot" composition is set at C:H:O = 100:78:17 in atomic ratio.  We estimated the density of soot at the reference condition of 1 bar and 298 K by examining the correlation between the density and mean atomic number for a diverse range of carbon-bearing phases, without considering its thermodynamic stability.  The thermoelastic properties of soot are bounded by the highly compressible water ice and the highly incompressible diamond. These simplifying assumptions allow us to bound the M-R relations of soot planets, for comparisons with exoplanets with well-measured masses and radii (defined here as uncertainties less than 20\% and 10\%, respectively).

We consider two end-member model planets that are either fully stratified or fully mixed. A multi-layer stratified planet consists of a metallic core enclosed in a silicate mantle, and where present, overlain by a hydrocarbon-rich layer, and finally a water ice surface layer.  A single-layer mixed planet represents an extreme case where iron, silicate, soot, and ice are fully mixed at all depths. It is usually assumed that planets have sufficient energy to differentiate and form layered interiors. However, the high temperatures prevailing in sub-Neptunes may produce exotic chemistry in which metal, silicate and water are fully miscible\citep{Young2025arXiv250700947Y, young2024phase}. We expect real planets to lie in between these two extremes, because water, soot, silicate, and metal may be partially or wholly miscible. Details of the mass-radius (M-R) calculation are described in the Appendix.
%Mass-radius relations place important constraints on the bulk composition of planets. 
%The 1 bar density of each phase is a key EoS parameter for computing mass–radius relations. 
%As the fate of soot in planets and the relevant phases at applicable conditions are unknown,
%However,  there also exists a similar gradient in the solid-state carbon content of minor solar system bodies \citep{Bergin15},  carried by macromolecular organics \citep{Alexander17}. There is one key distinction; the mass fraction of carbon of bodies in the inner solar system (e.g., meteorites/asteroids) is significantly higher than the water content \citep{Hirschmann16, Broadley2022}.  In comet 67P, the best characterized object that formed beyond the snow line, organics comprise nearly 45\%of the estimated refractory mass \citep{Bardyn2017}.

\section{Results and Discussion}
Our composition models (Table~\ref{Tab:composition}) integrated astronomical and cosmochemical constraints on solar composition \citep{Lodders03}, the disposition of carbon during planet formation \citep{Li21}, compositions of insoluble organic matter (IOM) in chondrites \citep{Alexander17}, and the dust-to-ice ratios in cometary materials \citep{Rubin19}.   

\begin{figure}[h]
\centering
\includegraphics [width=\linewidth]{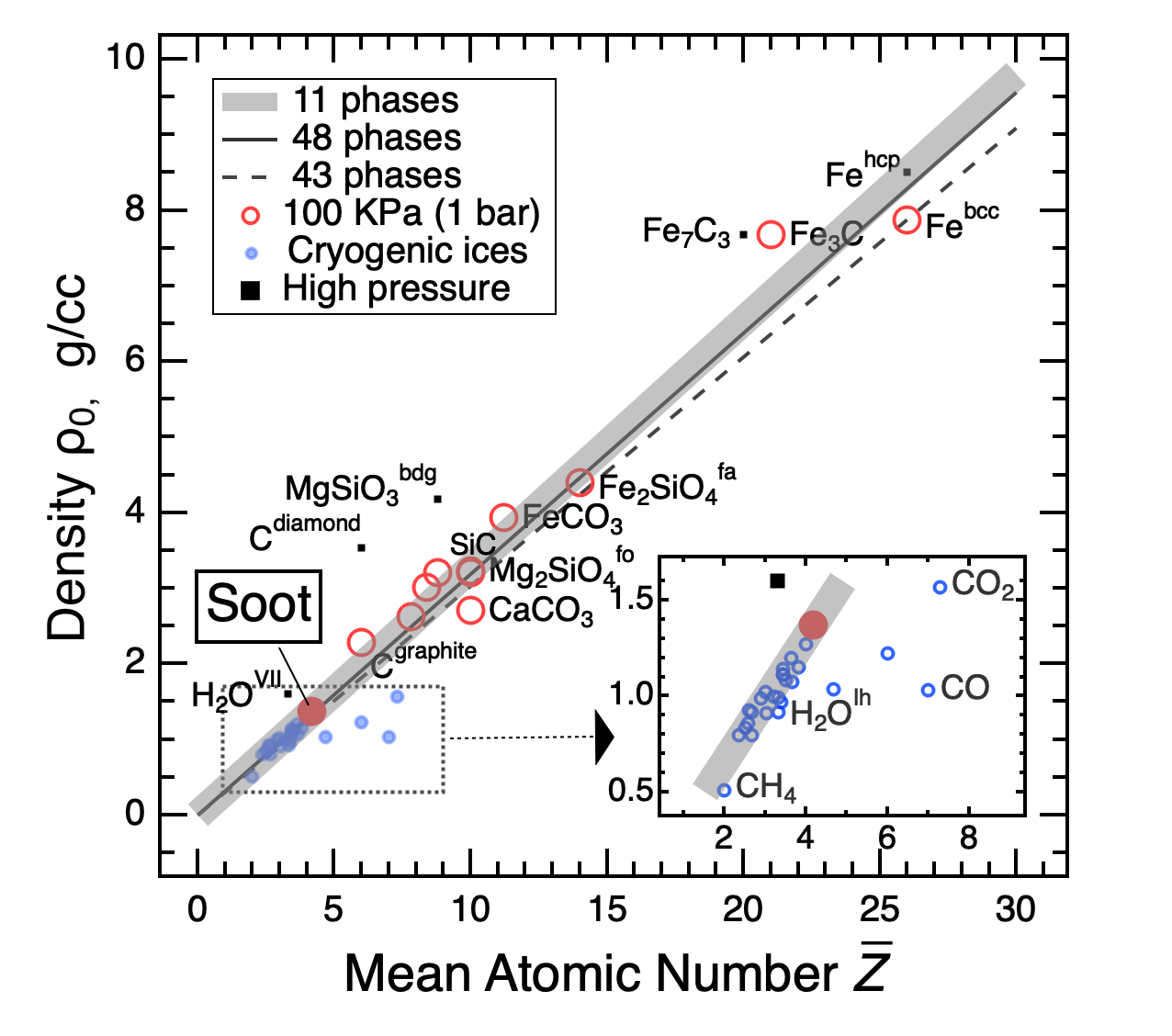}
%\captionsetup{justification=justified}
\caption{Correlation between density at 100 kPa (1 bar) $\rho_0$ and mean atomic number $\overline{Z}$ for a range of plausible solid phases in rocky or icy planetary interiors (data sources in Table~\ref{Tab:rho_z}). The gray band is a linear fit, $\rho_0$ = 0.327 $\overline{Z}$, for 11 selected phases that are thermodynamically stable at 100 kPa and 300 K (red open circles). Measured or fitted 100 kPa densities of high-pressure phases (black filled squares) mostly plot above the gray band, as expected. \hl{The cryogenic ice phases (blue open circles) may be considerably less dense and plot below the gray band.} The solid line is linear fit, $\rho_0$ = 0.318 $\overline{Z}$, for 43 phases that are thermodynamically stable at 100 kPa and 300 K. The dashed line is linear fit, $\rho_0$ = 0.303 $\overline{Z}$, for all 48 phases. \hl{Fitting the 43 phases and all 48 phases yielded 1.26 g/cc and 1.33 g/cc for the fictitious soot component, respectively.} The arrow points to soot (the red filled circle, C:H:O = 100:78:17 in atomic ratio, mean atomic number = 4.17), with an estimated density at 100 kbar and 300 K, $\rho_0$, of 1.32(6) g/cc.}
\label{fig:rho_z}
\end{figure}

 The densities and mean atomic numbers of a diverse range of carbon-bearing and planet-forming phases are compiled and examined (Table~\ref{Tab:rho_z}). We found that the correlation between 1 bar densities $\rho_0$ and mean atomic number $\overline{Z}$ is nearly linear (Fig.~\ref{fig:rho_z}), suggesting that bonding and structural arrangement have secondary influences on the densities, and therefore they can be ignored in estimating the 1 bar densities. With C:H:O = 100:78:17 in atomic ratio, the $\overline{Z}$ of the soot component is 4.17. Applying the fitted relation $\rho_0$ = 0.317(15) $\overline{Z}$, the density of the soot component is estimated at $\rho_0$ of 1.32(6) g/cc at 100 kPa and 300 K.  This estimate is supported by Bayesian analysis of the correlation (Fig.~\ref{fig:bayesian}). 

\clearpage
 
\begin{figure}[h]
\centering
    \begin{subfigure}{0.5\textwidth}
        \centering
        \includegraphics[width=0.9\linewidth]{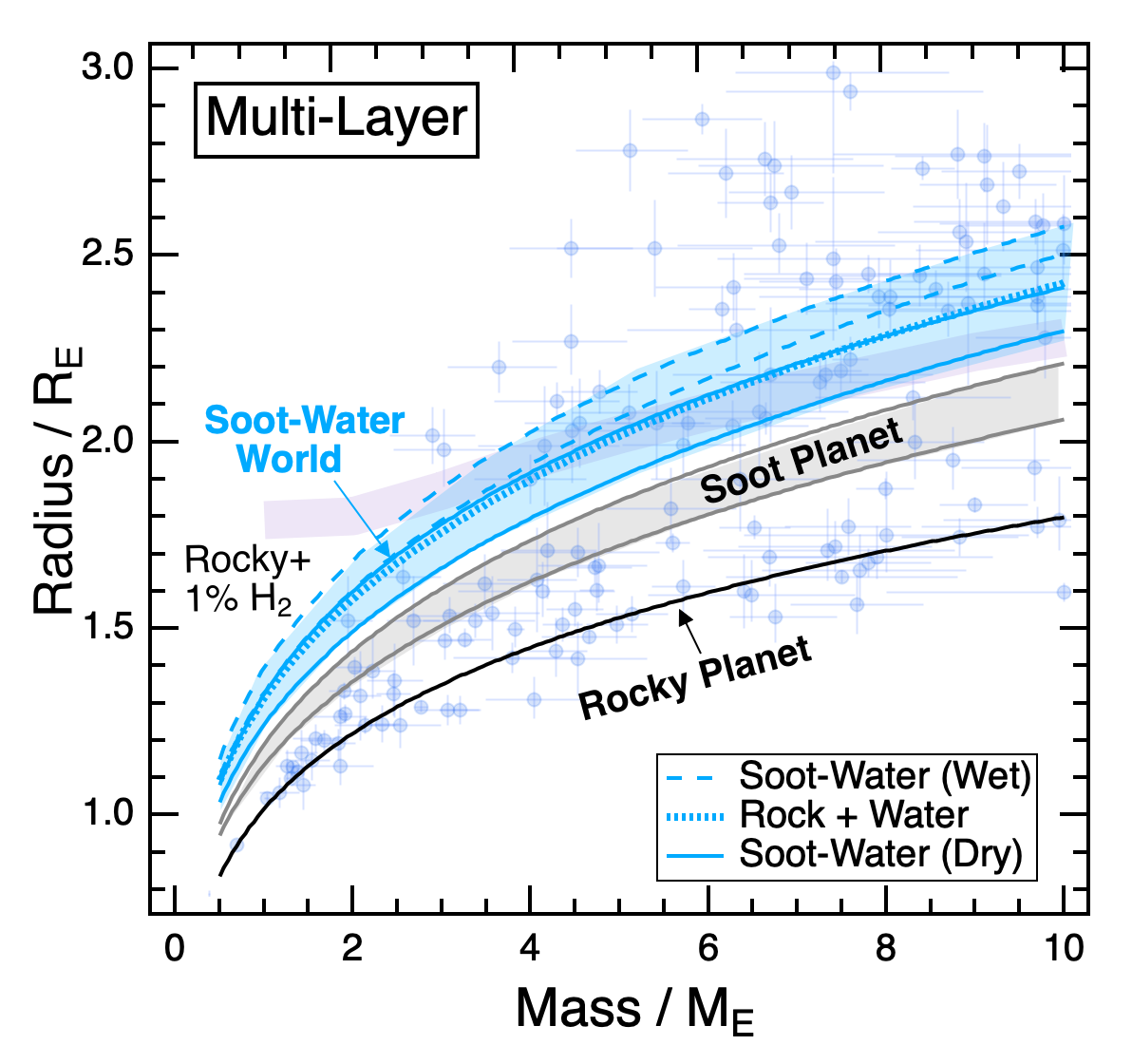}
    \end{subfigure}
    \hskip -0.3in
%\hfill
     \begin{subfigure}{0.5\textwidth}
         \centering
         \includegraphics[width=0.9\linewidth]{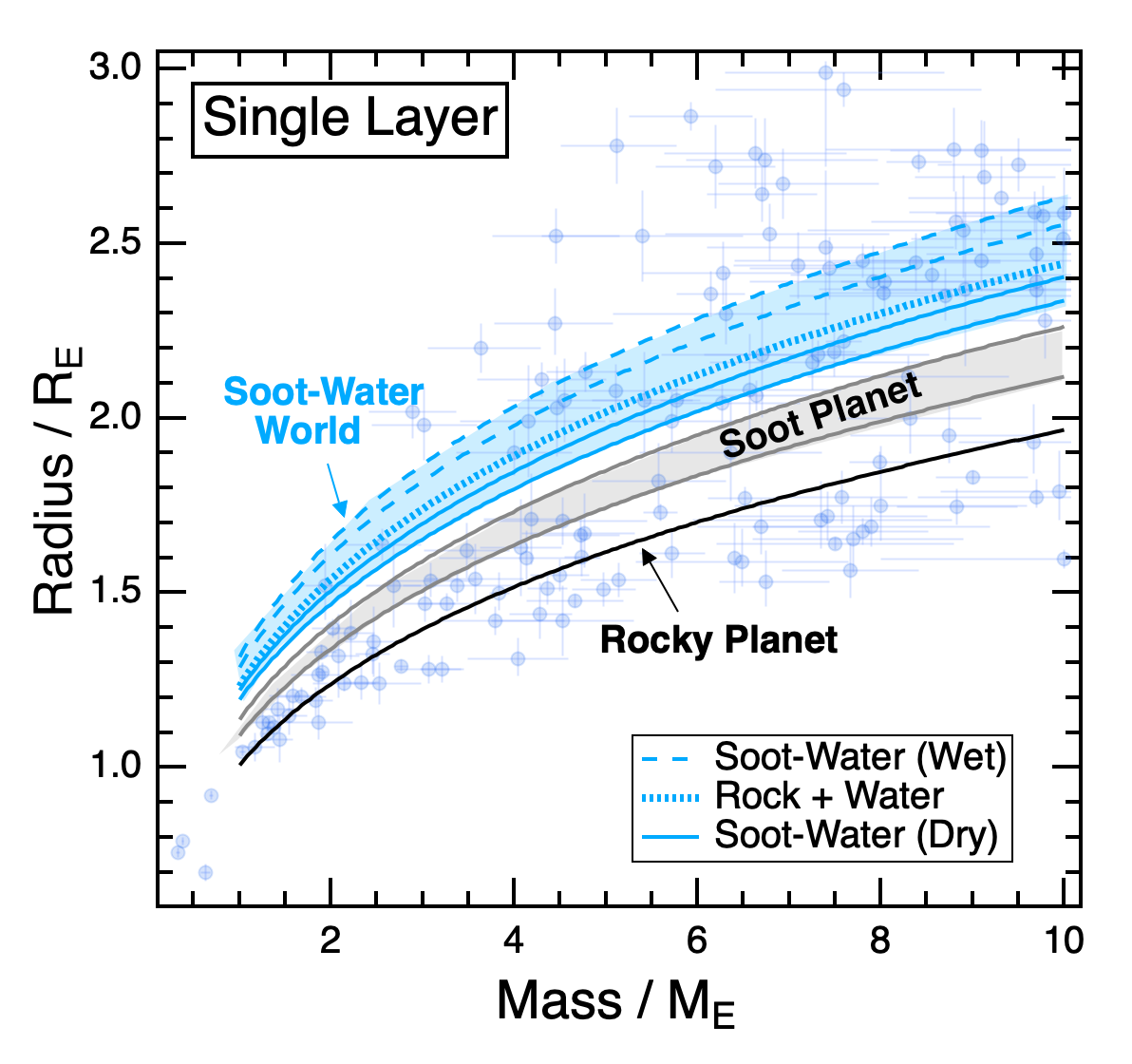}
    \end{subfigure}

\caption{Mass-Radius relations for model Earth-like rocky planets (black curves), soot planets (gray bands), and soot-water worlds (blue bands). Panel A is for multi-layer planets and Panel B is for single-layer planets. The shaded bands for the soot planets and soot-water worlds encompass end-member cases where the soot component varies from highly incompressible (diamond) to highly compressible (water ice) (Fig.~\ref{fig:eos}). Shown for comparison are curves for model planets consisting of 50\% Earth-like material and 50\% water ice (blue dotted curves) along with Earth-like rock planets with 1\% by mass H$_2$ envelope (purple band), as well as exoplanets with up to 3 Earth radii and up to 10 Earth masses, and for which masses and radii are known to $<$20\% and $<$10\%, respectively (blue crosses, NASA Exoplanet Archive: \url{https://exoplanetarchive.ipac.caltech.edu}.)}
\label{Fig:m_r}
\end{figure}

The mass-radius relations of the model planets are established and compared to the collection of known exoplanets (Fig.~\ref{Fig:m_r}). For the model soot planet and soot-water worlds, a range of radii are possible at a given mass, reflecting the limits we assume on whether the soot component is highly incompressible as diamond or highly compressible as H$_{2}$O ice (Table~\ref{Tab:eos}, Fig.~\ref{fig:eos}). At a given composition and mass, the calculated radius of a differentiated planet is found to be smaller than its undifferentiated counterpart. The difference would be at least partially offset by thermal expansion associated with differentiation. Exoplanets of $\lesssim$10 Earth masses exhibit a wide range of densities, with many observed mini-Neptunes following similar mass-radius relationships as soot planets or soot-water worlds.  The predicted M-R relations for the realistic water worlds, which incorporates soot, is similar to that predicted previously for a 50\% water planet with no carbon \cite{Luque22}. Thus, based on the expected M-R relations alone, observed exoplanets previously interpreted as rock+water planets cannot be distinguished from water-rich planets incorporating significant soot. This suggests that a major compositional component of such planets may be overlooked. 

Previous studies also considered carbon-bearing planets, but they arise from different implied formation scenarios that do not account for refractory carbon in the protoplanetary disk, and therefore they differ from our soot planets in fundamental aspects. For example, the vast majority of stellar C/O ratios for known planetary systems mostly fall between 0.3 and 0.8, with the medium value at the solar value of 0.5 (Fig.~\ref{fig:star_c_o}). Carbon-rich exoplanets have been proposed for stars with C/O ratios that are close to or exceed unity. (e.g.,\citep{Madhu12, 2018JGRE..123.2295M, Hakim19}), In contrast, our model soot planets have sub-solar C/O ratios ranging from one-half to one-sixth of the solar value. Moreover, other studies that considered carbon-rich exoplanet \emph{atmospheres} have typically invoked ice-line arguments as the source of the augmented carbon, whereas our soot-rich planets formed over a wide range of the accretion disc beyond the soot line and contain a substantial hydrocarbon-rich component \citep[e.g.,][]{madhus11, oberg11_c2d, alidib14}. Recently, \citet{Peng2024ApJ...976..202P} explored the potential for carbon-rich mantles leading to the creation of Venus analogs with thick, carbon-dominated atmospheres. They adopted the CI level of carbon content for planets forming inside the water ice line and did not discuss the source of carbon in the formation process.  

The M-R relations shown in Fig.~\ref{Fig:m_r} have substantial degeneracy \citep[e.g.,][]{Seager2007, rogers10}. \hl{In this work, we identify new degeneracies that should be considered and that challenge the water world interpretation. Soot-rich planets or cores of sub-Neptunes (with or without substantive water inventories) are predicted by planet formation models that explain key astronomical and cosmochemical observations. They also adequately explain the observed mass-radius relations of low-mass low-density exoplanets. Therefore, we must contend with this level of compositional complexity, especially since the mass fractions of soot and water may vary depending on the their formation path} \citep{Binkert2023}. The model planets illustrated here are based on the range of observed dust to ice ratios in cometary material \citep{Rubin19}. This ratio may vary in part because a planet may lose volatiles during accretion via parent-body processing \citep{Lichtenberg2023} and impact degassing \citep{2018SSRv..214...34S}. 

Our model planets do not contain gaseous (H$_2$+He) envelopes. With increasing mass, planets are more likely to attract and retain appreciable nebular hydrogen \citep{Owen2013}. Therefore, H$_2$-rich atmospheres are quite likely present in massive planets, but are less likely to be preserved in low mass planets.  For more massive planets, a gaseous envelope may contribute to the mass - radius relations of observed low mean density planets. For example, a nebular envelope of 1\% planet mass increases the radius of a rocky planet by 20 to 30\% (Fig.~\ref{Fig:m_r}).  Therefore, relatively massive ($\gtrsim 1$ $M_\oplus$; \citep{owen2020hydrogen}) soot planets or soot-water worlds cannot be distinguished from rocky planets with H$_2$ envelopes based on M-R relationships alone. Nevertheless, we expect larger planets with hydrogen envelopes to have soot-rich mantles that will likely alter the atmospheric composition \citep{Bergin2023}.  Smaller planets and those close to their star are less likely to have retained their primary gaseous envelopes \citep{owen2020hydrogen}, so soot or soot-water worlds are more likely explanations for these low mass, low density planets.  \hl{A key point is that no single compositional scenario describes the full set of observed exoplanets, so considerable compositional diversity (i.e., different admixtures of iron, rock, water, hydrogen, \emph{and soot}) are needed to explain the full range of planetary radii observed at a given mass, with massive hydrogen envelopes being more likely at higher masses and above the so-called `radius valley'} \citep{fulton17}.

The distribution of carbonaceous phases in a soot-rich planet has been little-explored. Major possible sinks for carbon in known planets from the Solar System include the metallic cores of differentiated terrestrial planets \cite{Fischer20,Li21} and clathrate ice layers in the ice giants Uranus and Neptune \citep{Malamud2024}.  Metallic cores of planets can only store a small fraction of carbon delivered as soot: The maximum concentration of carbon in an iron-rich core is below 10 wt\% \cite{Dasgupta_Walker_2008}, corresponding to 2.5 wt.\% C in a planet with 25 wt.\% metallic iron, which is much less than the average C concentration of a soot planet (Table~\ref{Tab:composition}). Clathrate ices are plausible for planets with cold interiors \cite{Lunine_Stevenson_1985}, but not for those more thermally vigorous.

Soot-rich planets are expected to possess distinct atmospheric compositions. Thermal processing of soot-rich components in planetary interiors, either in isolated layers or mixed with silicate, is expected to yield refractory phases such as graphite, diamond, or carbide \citep{Davydov2004}.  =Through interactions with H$_2$-rich envelopes, the mantles would be reduced with high H/C ratios. Because the solubility of C-H-rich components in silicate liquid and minerals is limited \citep{Armstrong15, Ardia13}, significant fractions of methane and other simple hydrocarbons are expected to be released from the interior. A resulting methane-rich atmosphere may naturally lead to the formation of hydrocarbon hazes, akin to the tholins in Titan's atmosphere \citep{millerricci12,morley13,Bergin2023}.  Indeed, many sub-Neptunes have produced featureless spectra indicating the likely presence of clouds or photochemical haze, perhaps supporting our concept of a soot planet \citep[e.g.,][]{Kreidberg14, brande24}.  

%In our framework of a soot planet, the haze would naturally arise in the accompanying carbon-rich outer envelope.

%but other outcomes are feasible depending on planetary mass and temperature.

 %Hydrocarbon haze has been invoked as a likely cause \citep{millerricci12,morley13}, which would naturally arise in methane-rich atmospheres consistent with the soot planet scenario \cite{Bergin2023}.

%Outgassing of water-poor soot planets would naturally yield the ingredients for haze production when exposed to stellar UV photons in the upper atmosphere, possibly accounting for featureless transmission spectra that appear common in the exoplanetary inventory. 

%\hl{Among the sub-Neptunes with published atmospheric spectra, TOI-270\,d stands out as being qualitatively consistent with the predicted composition of a soot planet, thus motivating further scrutiny of this type of planet and future modeling work to flesh out predictions for atmospheric composition of soot planets with accreted primary atmospheres.}

JWST observations of sub-Neptunes have recently allowed for the direct detection of carbon-bearing species in their atmospheres \citep{madhu23,holmberg24,benneke24}.  An active area of research attempts to tie the observed atmospheric composition to the properties of a planet's bulk interior \citep[e.g.,][]{nixon24,wogan24,rigby24}.  Presently it is unclear whether the observed composition of any individual sub-Neptune atmosphere indicates a significant fraction of carbon in its bulk, but tentative evidence supports our definition of either a soot planet or a soot-water world. For example, the discovery of abundant CO$_2$ and CH$_4$ in the atmospheres of K2-18\,b and TOI-270\,d, accompanied by H$_2$O in the latter case \citep{madhu23, holmberg24, benneke24}, indicates a reservoir of carbon participating in atmospheric chemistry.  Linking the atmospheric carbon-to-oxygen ratio (C/O) to the carbon content of the planet's bulk interior requires models of the interface between the deep atmosphere and an underlying magma ocean, which will be the subject of future work. In the absence of a considerable soot component, one would qualitatively expect a very low atmospheric C/O because water outgasses readily from silicates in contact with hydrogen-dominated atmospheres, but we are not aware of a comparable mechanism to release abundant carbon-rich gases from an initially carbon-poor interior \citep{blanchard22,seo24}. Thus the moderately high C/O for the sub-Neptune TOI-270\,d ($\sim$0.5) may indicate that it is a soot planet. 

\section{Implications}
Some characteristics of the interiors of soot-rich planets may inhibit habitability. Abundant diamond in a solidified silicate mantle could increase interior viscosity and thermal conductivity by several orders of magnitude, thus retarding mantle convection and volatile cycling, and potentially yielding inhospitable surface conditions \citep{Unterborn2014}. A carbon-rich core may not provide the buoyancy needed to support a global magnetic field, which protects life from stellar wind and cosmic radiation, though other dynamo mechanisms are possible in a sub-Neptune that retains a partially molten deep silicate interior \citep{Stixrude2021}. 

Alternately, soot-rich planets can have favorable consequences for habitability and the development of life.  Persistent supplies of methane and other reduced gases, including likely significant partial pressures of NH$_3$ and H$_2$ \cite{liggins2022}, are anticipated as volcanogenic products from the thermally processed interiors of soot-rich planets \citep{Bergin2023}. Abundant methane with reducing atmospheric and oceanic conditions are possibly a prerequisite for progression of prebiotic chemistry \citep{McCollom13}. An instructive analog may be found on Titan, where hydrolysis of methane-produced hazes have been shown via laboratory experiments to form simple amino acids \citep{Cleaves2014}.  Although reactions between organics and liquid water may not occur on Titan \citep{Neish2024}, organics could be created in the atmosphere of soot-water worlds and provide a consistent source to potentially foster a biosphere in underlying oceans.
\\
\\
\noindent Correspondence should be addressed to Jie Li (jackieli@umich.edu) and Edwin Bergin (ebergin@umich.edu). \\[0.25cm]

\textbf{Author contributions} Jie Li compiled data and performed the calculations. All authors then co-wrote the manuscript and agreed on the conclusions. 

\textbf{Acknowledgment} 
The research shown here acknowledges use of the Hypatia Catalog Database, an online compilation of stellar abundance data as described in \citet{Hypatia}, which was supported by NASA's Nexus for Exoplanet System Science (NExSS) research coordination network and the Vanderbilt Initiative in Data-Intensive Astrophysics (VIDA).
JL acknowledges support from National Science Foundation (NSF) grant EAR2317024. MMH acknowledges support from NSF grant EAR2246915 and National Aeronautics and Space Administration (NASA) grant 80NSSC19K0959. EAB and GAB acknowledge support from NASA grant 80NSSC24K0149. EM-RK acknowledges funding from AEThER via the Alfred P. Sloan Foundation under grant G202114194. FJC and EAB acknowledge support from NASA Award 80NSSC20K0259.  This project is supported, in part, by funding from Two Sigma Investments, LP to EAB. Any opinions, findings, and conclusions or recommendations expressed in this material are those of the authors and do not necessarily reflects the views of Two Sigma Investments, LP.

%\textbf{Competing interest} The authors declare no competing interests.

%\textbf{Data Availability Statement} The authors declare that the data supporting the findings of this study are available in the article, its references and its supplementary information files.

\textbf{Code Availability Statement}  The codes for calculating the M-R relations are made available on Zenodo \citep{li25Zenodo}.
\newpage
% https://journals.aas.org/aastexguide/#softwareandthirdparty

\bibliography{z}

@MISC{li25Zenodo,
author       = {{Li}, Jie},
title        = "{{Soot Planet Code}}",
month        = july,
year         = 2025,
doi          = {10.5281/zenodo.16101631},
version      = {1.0},
publisher    = {Zenodo},
url          = {https://doi.org/10.5281/zenodo.16101631}
}

@ARTICLE{Woodward2021,
       author = {{Woodward}, Charles E. and {Wooden}, Diane H. and {Harker}, David E. and {Kelley}, Michael S.~P. and {Russell}, Ray W. and {Kim}, Daryl L.},
        title = "{The Coma Dust of Comet C/2013 US$_{10}$ (Catalina): A Window into Carbon in the Solar System}",
      journal = {\psj},
         year = 2021,
        month = feb,
       volume = {2},
       number = {1},
          eid = {25},
        pages = {25},
          doi = {10.3847/PSJ/abca3e},
archivePrefix = {arXiv},
       eprint = {2011.06943},
 primaryClass = {astro-ph.EP},
       adsurl = {https://ui.adsabs.harvard.edu/abs/2021PSJ.....2...25W}
}

@ARTICLE{Fomenkova1999,
       author = {{Fomenkova}, Marina N.},
        title = "{On the Organic Refractory Component of Cometary Dust}",
      journal = {\ssr},
         year = 1999,
        month = oct,
       volume = {90},
        pages = {109-114},
          doi = {10.1023/A:1005237828783},
       adsurl = {https://ui.adsabs.harvard.edu/abs/1999SSRv...90..109F}
}

@ARTICLE{2024A&A...691A.352W,
       author = {{Williams}, J.~T. and {G{\"a}nsicke}, B.~T. and {Swan}, A. and {O'Brien}, M.~W. and {Izquierdo}, P. and {Cutolo}, A.-M. and {Cunningham}, T.},
        title = "{PEWDD: A database of white dwarfs enriched by exo-planetary material}",
      journal = {\aap},
         year = 2024,
        month = nov,
       volume = {691},
          eid = {A352},
        pages = {A352},
          doi = {10.1051/0004-6361/202450509},
archivePrefix = {arXiv},
       eprint = {2409.16046},
 primaryClass = {astro-ph.EP},
       adsurl = {https://ui.adsabs.harvard.edu/abs/2024A&A...691A.352W}
}

@ARTICLE{koester2014atmospheric,
       author = {{Koester}, D. and {Provencal}, J. and {G{\"a}nsicke}, B.~T.},
        title = "{Atmospheric parameters and carbon abundance for hot DB white dwarfs}",
      journal = {\aap},
         year = 2014,
        month = aug,
       volume = {568},
          eid = {A118},
        pages = {A118},
          doi = {10.1051/0004-6361/201424231},
archivePrefix = {arXiv},
       eprint = {1407.6157},
 primaryClass = {astro-ph.SR},
       adsurl = {https://ui.adsabs.harvard.edu/abs/2014A&A...568A.118K}
}

@ARTICLE{Shieh06ppv,
       author = {{Shieh}, Sean R. and {Duffy}, Thomas S. and {Kubo}, Atsushi and {Shen}, Guoyin and {Prakapenka}, Vitali B. and {Sata}, Nagayoshi and {Hirose}, Kei and {Ohishi}, Yasuo},
        title = "{Equation of state of the postperovskite phase synthesized from a natural (Mg,Fe)SiO$_{3}$ orthopyroxene}",
      journal = {Proceedings of the National Academy of Science},
         year = 2006,
        month = feb,
       volume = {103},
       number = {9},
        pages = {3039-3043},
          doi = {10.1073/pnas.0506811103},
       adsurl = {https://ui.adsabs.harvard.edu/abs/2006PNAS..103.3039S}
}

@ARTICLE{Finkelstein14Forsterite,
       author = {{Finkelstein}, G.~J. and {Dera}, P.~K. and {Jahn}, S. and {Oganov}, A.~R. and {Holl}, C.~M. and {Meng}, Y. and {Duffy}, T.~S.},
        title = "{Phase transitions and equation of state of forsterite to 90 GPa from single-crystal X-ray diffraction and molecular modeling}",
      journal = {American Mineralogist},
         year = 2014,
        month = jan,
       volume = {99},
       number = {1},
        pages = {35-43},
          doi = {10.2138/am.2014.4526},
       adsurl = {https://ui.adsabs.harvard.edu/abs/2014AmMin..99...35F}
}

@ARTICLE{Wang12Graphite, 
       author = {{Wang}, Yuejian and {Panzik}, Joseph E. and {Kiefer}, Boris and {Lee}, Kanani K.~M.},
        title = "{Crystal structure of graphite under room-temperature compression and decompression}",
      journal = {Scientific Reports},
         year = 2012,
        month = jul,
       volume = {2},
          eid = {520},
        pages = {520},
          doi = {10.1038/srep00520},
       adsurl = {https://ui.adsabs.harvard.edu/abs/2012NatSR...2..520W}
}

@ARTICLE{2018SSRv..214...34S,
       author = {{Schlichting}, Hilke E. and {Mukhopadhyay}, Sujoy},
        title = "{Atmosphere Impact Losses}",
      journal = {\ssr},
         year = 2018,
        month = feb,
       volume = {214},
       number = {1},
          eid = {34},
        pages = {34},
          doi = {10.1007/s11214-018-0471-z},
       adsurl = {https://ui.adsabs.harvard.edu/abs/2018SSRv..214...34S}
}

@ARTICLE{brande24,
       author = {{Brande}, Jonathan and {Crossfield}, Ian J.~M. and {Kreidberg}, Laura and {Morley}, Caroline V. and {Barman}, Travis and {Benneke}, Bj{\"o}rn and {Christiansen}, Jessie L. and {Dragomir}, Diana and {Fortney}, Jonathan J. and {Greene}, Thomas P. and {Hardegree-Ullman}, Kevin K. and {Howard}, Andrew W. and {Knutson}, Heather A. and {Lothringer}, Joshua D. and {Mikal-Evans}, Thomas},
        title = "{Clouds and Clarity: Revisiting Atmospheric Feature Trends in Neptune-size Exoplanets}",
      journal = {\apjl},
         year = 2024,
        month = jan,
       volume = {961},
       number = {1},
          eid = {L23},
        pages = {L23},
          doi = {10.3847/2041-8213/ad1b5c},
archivePrefix = {arXiv},
       eprint = {2310.07714},
 primaryClass = {astro-ph.EP},
       adsurl = {https://ui.adsabs.harvard.edu/abs/2024ApJ...961L..23B}
}

@ARTICLE{alidib14,
       author = {{Ali-Dib}, Mohamad and {Mousis}, Olivier and {Petit}, Jean-Marc and {Lunine}, Jonathan I.},
        title = "{Carbon-rich Planet Formation in a Solar Composition Disk}",
      journal = {\apj},
         year = 2014,
        month = apr,
       volume = {785},
       number = {2},
          eid = {125},
        pages = {125},
          doi = {10.1088/0004-637X/785/2/125},
archivePrefix = {arXiv},
       eprint = {1402.5182},
 primaryClass = {astro-ph.EP},
       adsurl = {https://ui.adsabs.harvard.edu/abs/2014ApJ...785..125A}
}

@ARTICLE{Hirschmann2021PNAS..11826779H,
       author = {{Hirschmann}, Marc M. and {Bergin}, Edwin A. and {Blake}, Geoff A. and {Ciesla}, Fred J. and {Li}, Jie},
        title = "{Early volatile depletion on planetesimals inferred from C{\textendash}S systematics of iron meteorite parent bodies}",
      journal = {Proceedings of the National Academy of Science},
         year = 2021,
        month = mar,
       volume = {118},
       number = {13},
          eid = {e2026779118},
        pages = {e2026779118},
          doi = {10.1073/pnas.2026779118},
archivePrefix = {arXiv},
       eprint = {2104.02706},
 primaryClass = {astro-ph.EP},
       adsurl = {https://ui.adsabs.harvard.edu/abs/2021PNAS..11826779H}
}

@ARTICLE{Ciesla2005E&PSL.231....1C,
       author = {{Ciesla}, Fred and {Lauretta}, Dante},
        title = "{Radial migration and dehydration of phyllosilicates in the solar nebula}",
      journal = {Earth and Planetary Science Letters},
         year = 2005,
        month = feb,
       volume = {231},
       number = {1-2},
        pages = {1-8},
          doi = {10.1016/j.epsl.2004.12.022},
       adsurl = {https://ui.adsabs.harvard.edu/abs/2005E&PSL.231....1C}
}

@ARTICLE{Peng2024ApJ...976..202P,
       author = {{Peng}, Bo and {Valencia}, Diana},
        title = "{Puffy Venuses: The Mass{\textendash}Radius Impact of Carbon-rich Atmospheres on Lava Worlds}",
      journal = {\apj},
         year = 2024,
        month = dec,
       volume = {976},
       number = {2},
          eid = {202},
        pages = {202},
          doi = {10.3847/1538-4357/ad6f03},
archivePrefix = {arXiv},
       eprint = {2405.08998},
 primaryClass = {astro-ph.EP},
       adsurl = {https://ui.adsabs.harvard.edu/abs/2024ApJ...976..202P}
}

@ARTICLE{Young2025arXiv250700947Y,
       author = {{Young}, Edward D. and {Werlen}, Aaron and {Marcum}, Sarah P. and {Dullemond}, Cornelis P.},
        title = "{Differentiation, the exception not the rule -- Evidence for full miscibility in sub-Neptune interiors}",
      journal = {arXiv e-prints},
         year = 2025,
        month = jul,
          eid = {arXiv:2507.00947},
        pages = {arXiv:2507.00947},
          doi = {10.48550/arXiv.2507.00947},
archivePrefix = {arXiv},
       eprint = {2507.00947},
 primaryClass = {astro-ph.EP},
       adsurl = {https://ui.adsabs.harvard.edu/abs/2025arXiv250700947Y}
}

@ARTICLE{Minissale22,
       author = {{Minissale}, Marco and {Aikawa}, Yuri and {Bergin}, Edwin and {Bertin}, Mathieu and {Brown}, Wendy A. and {Cazaux}, Stephanie and {Charnley}, Steven B. and {Coutens}, Audrey and {Cuppen}, Herma M. and {Guzman}, Victoria and {Linnartz}, Harold and {McCoustra}, Martin R.~S. and {Rimola}, Albert and {Schrauwen}, Johanna G.~M. and {Toubin}, Celine and {Ugliengo}, Piero and {Watanabe}, Naoki and {Wakelam}, Valentine and {Dulieu}, Francois},
        title = "{Thermal Desorption of Interstellar Ices: A Review on the Controlling Parameters and Their Implications from Snowlines to Chemical Complexity}",
      journal = {ACS Earth and Space Chemistry},
         year = 2022,
        month = mar,
       volume = {6},
       number = {3},
        pages = {597-630},
          doi = {10.1021/acsearthspacechem.1c00357},
archivePrefix = {arXiv},
       eprint = {2201.07512},
 primaryClass = {astro-ph.GA},
       adsurl = {https://ui.adsabs.harvard.edu/abs/2022ESC.....6..597M}
}

@ARTICLE{Prakapenka21,
       author = {{Prakapenka}, Vitali B. and {Holtgrewe}, Nicholas and {Lobanov}, Sergey S. and {Goncharov}, Alexander F.},
        title = "{Structure and properties of two superionic ice phases}",
      journal = {Nature Physics},
         year = 2021,
        month = oct,
       volume = {17},
       number = {11},
        pages = {1233-1238},
          doi = {10.1038/s41567-021-01351-8},
       adsurl = {https://ui.adsabs.harvard.edu/abs/2021NatPh..17.1233P}
}

@ARTICLE{White25,
       author = {{White}, Nathaniel I. and {Li}, Jie},
        title = "{Initial Thermal States of Super-Earth Exoplanets and Implications for Early Dynamos}",
      journal = {Journal of Geophysical Research (Planets)},
         year = 2025,
        month = feb,
       volume = {130},
       number = {2},
        pages = {2024JE008550},
          doi = {10.1029/2024JE008550},
       adsurl = {https://ui.adsabs.harvard.edu/abs/2025JGRE..13008550W}
}

@ARTICLE{2025NatAs...9..199G,
       author = {{Glavin}, Daniel P. and {Dworkin}, Jason P. and {Alexander}, Conel M. O'D. and {Aponte}, Jos{\'e} C. and {Baczynski}, Allison A. and {Barnes}, Jessica J. and {Bechtel}, Hans A. and {Berger}, Eve L. and {Burton}, Aaron S. and {Caselli}, Paola and {Chung}, Angela H. and {Clemett}, Simon J. and {Cody}, George D. and {Dominguez}, Gerardo and {Elsila}, Jamie E. and {Farnsworth}, Kendra K. and {Foustoukos}, Dionysis I. and {Freeman}, Katherine H. and {Furukawa}, Yoshihiro and {Gainsforth}, Zack and {Graham}, Heather V. and {Grassi}, Tommaso and {Giuliano}, Barbara Michela and {Hamilton}, Victoria E. and {Haenecour}, Pierre and {Heck}, Philipp R. and {Hofmann}, Amy E. and {House}, Christopher H. and {Huang}, Yongsong and {Kaplan}, Hannah H. and {Keller}, Lindsay P. and {Kim}, Bumsoo and {Koga}, Toshiki and {Liss}, Michael and {McLain}, Hannah L. and {Marcus}, Matthew A. and {Matney}, Mila and {McCoy}, Timothy J. and {McIntosh}, Oph{\'e}lie M. and {Mojarro}, Angel and {Naraoka}, Hiroshi and {Nguyen}, Ann N. and {Nuevo}, Michel and {Nuth}, Joseph A. and {Oba}, Yasuhiro and {Parker}, Eric T. and {Peretyazhko}, Tanya S. and {Sandford}, Scott A. and {Santos}, Ewerton and {Schmitt-Kopplin}, Philippe and {Seguin}, Frederic and {Simkus}, Danielle N. and {Shahid}, Anique and {Takano}, Yoshinori and {Thomas-Keprta}, Kathie L. and {Tripathi}, Havishk and {Weiss}, Gabriella and {Zheng}, Yuke and {Lunning}, Nicole G. and {Righter}, Kevin and {Connolly}, Harold C. and {Lauretta}, Dante S.},
        title = "{Abundant ammonia and nitrogen-rich soluble organic matter in samples from asteroid (101955) Bennu}",
      journal = {Nature Astronomy},
         year = 2025,
        month = feb,
       volume = {9},
        pages = {199-210},
          doi = {10.1038/s41550-024-02472-9},
       adsurl = {https://ui.adsabs.harvard.edu/abs/2025NatAs...9..199G}
}

@ARTICLE{2025GeocJ..59...45Y,
       author = {{Yokoyama}, Tetsuya and {Dauphas}, Nicolas and {Fukai}, Ryota and {Usui}, Tomohiro and {Tachibana}, Shogo and {Sch{\"o}nb{\"a}chler}, Maria and {Busemann}, Henner and {Abe}, Masanao and {Yada}, Toru},
        title = "{The elemental abundances of Ryugu: Assessment of chemical heterogeneities and the nugget effect}",
      journal = {Geochemical Journal},
         year = 2025,
        month = jan,
       volume = {59},
       number = {2},
        pages = {45-63},
          doi = {10.2343/geochemj.GJ25002},
archivePrefix = {arXiv},
       eprint = {2405.04500},
 primaryClass = {astro-ph.EP},
       adsurl = {https://ui.adsabs.harvard.edu/abs/2025GeocJ..59...45Y}
}

@ARTICLE{2019NatAs...3..332H,
       author = {{Hamilton}, V.~E. and {Simon}, A.~A. and {Christensen}, P.~R. and {Reuter}, D.~C. and {Clark}, B.~E. and {Barucci}, M.~A. and {Bowles}, N.~E. and {Boynton}, W.~V. and {Brucato}, J.~R. and {Cloutis}, E.~A. and {Connolly}, H.~C. and {Donaldson Hanna}, K.~L. and {Emery}, J.~P. and {Enos}, H.~L. and {Fornasier}, S. and {Haberle}, C.~W. and {Hanna}, R.~D. and {Howell}, E.~S. and {Kaplan}, H.~H. and {Keller}, L.~P. and {Lantz}, C. and {Li}, J. -Y. and {Lim}, L.~F. and {McCoy}, T.~J. and {Merlin}, F. and {Nolan}, M.~C. and {Praet}, A. and {Rozitis}, B. and {Sandford}, S.~A. and {Schrader}, D.~L. and {Thomas}, C.~A. and {Zou}, X. -D. and {Lauretta}, D.~S. and {Osiris-Rex Team}},
        title = "{Evidence for widespread hydrated minerals on asteroid (101955) Bennu}",
      journal = {Nature Astronomy},
         year = 2019,
        month = mar,
       volume = {3},
        pages = {332-340},
          doi = {10.1038/s41550-019-0722-2},
       adsurl = {https://ui.adsabs.harvard.edu/abs/2019NatAs...3..332H}
}

@ARTICLE{2015Icar..246...21M,
       author = {{Malamud}, Uri and {Prialnik}, Dina},
        title = "{Modeling Kuiper belt objects Charon, Orcus and Salacia by means of a new equation of state for porous icy bodies}",
      journal = {\icarus},
         year = 2015,
        month = jan,
       volume = {246},
        pages = {21-36},
          doi = {10.1016/j.icarus.2014.02.027},
       adsurl = {https://ui.adsabs.harvard.edu/abs/2015Icar..246...21M}
}

@ARTICLE{Valencia2007ApJ...665.1413V,
       author = {{Valencia}, Diana and {Sasselov}, Dimitar D. and {O'Connell}, Richard J.},
        title = "{Detailed Models of Super-Earths: How Well Can We Infer Bulk Properties?}",
      journal = {\apj},
         year = 2007,
        month = aug,
       volume = {665},
       number = {2},
        pages = {1413-1420},
          doi = {10.1086/519554},
archivePrefix = {arXiv},
       eprint = {0704.3454},
 primaryClass = {astro-ph},
       adsurl = {https://ui.adsabs.harvard.edu/abs/2007ApJ...665.1413V}
}

@ARTICLE{Sotin2007Icar..191..337S,
       author = {{Sotin}, C. and {Grasset}, O. and {Mocquet}, A.},
        title = "{Mass radius curve for extrasolar Earth-like planets and ocean planets}",
      journal = {\icarus},
         year = 2007,
        month = nov,
       volume = {191},
       number = {1},
        pages = {337-351},
          doi = {10.1016/j.icarus.2007.04.006},
       adsurl = {https://ui.adsabs.harvard.edu/abs/2007Icar..191..337S}
}

@ARTICLE{2019Icar..326...10B,
       author = {{Bierson}, C.~J. and {Nimmo}, F.},
        title = "{Using the density of Kuiper Belt Objects to constrain their composition and formation history}",
      journal = {\icarus},
         year = 2019,
        month = jul,
       volume = {326},
        pages = {10-17},
          doi = {10.1016/j.icarus.2019.01.027},
       adsurl = {https://ui.adsabs.harvard.edu/abs/2019Icar..326...10B}
}

@ARTICLE{2014ApJ...787...81M,
       author = {{Moriarty}, John and {Madhusudhan}, Nikku and {Fischer}, Debra},
        title = "{Chemistry in an Evolving Protoplanetary Disk: Effects on Terrestrial Planet Composition}",
      journal = {\apj},
         year = 2014,
        month = may,
       volume = {787},
       number = {1},
          eid = {81},
        pages = {81},
          doi = {10.1088/0004-637X/787/1/81},
archivePrefix = {arXiv},
       eprint = {1405.3253},
 primaryClass = {astro-ph.EP},
       adsurl = {https://ui.adsabs.harvard.edu/abs/2014ApJ...787...81M}
}

@ARTICLE{2025AJ....169..180S,
       author = {{Shakespeare}, Cody J. and {Li}, Min and {Huang}, Shichun and {Zhu}, Zhaohuan and {Steffen}, Jason H.},
        title = "{The Effects of the Carbon-to-oxygen Ratio on the Condensate Compositions around Solar-like Stars}",
      journal = {\aj},
         year = 2025,
        month = mar,
       volume = {169},
       number = {3},
          eid = {180},
        pages = {180},
          doi = {10.3847/1538-3881/adaead},
archivePrefix = {arXiv},
       eprint = {2408.07761},
 primaryClass = {astro-ph.EP},
       adsurl = {https://ui.adsabs.harvard.edu/abs/2025AJ....169..180S}
}

@ARTICLE{1994ApJ...421..615P,
       author = {{Pollack}, James B. and {Hollenbach}, David and {Beckwith}, Steven and {Simonelli}, Damon P. and {Roush}, Ted and {Fong}, Wesley},
        title = "{Composition and Radiative Properties of Grains in Molecular Clouds and Accretion Disks}",
      journal = {\apj},
         year = 1994,
        month = feb,
       volume = {421},
        pages = {615},
          doi = {10.1086/173677},
       adsurl = {https://ui.adsabs.harvard.edu/abs/1994ApJ...421..615P}
}

@ARTICLE{2005ApJ...631.1134A,
       author = {{Andrews}, Sean M. and {Williams}, Jonathan P.},
        title = "{Circumstellar Dust Disks in Taurus-Auriga: The Submillimeter Perspective}",
      journal = {\apj},
         year = 2005,
        month = oct,
       volume = {631},
       number = {2},
        pages = {1134-1160},
          doi = {10.1086/432712},
archivePrefix = {arXiv},
       eprint = {astro-ph/0506187},
 primaryClass = {astro-ph},
       adsurl = {https://ui.adsabs.harvard.edu/abs/2005ApJ...631.1134A}
}

@ARTICLE{2010A&A...512A..15R,
       author = {{Ricci}, L. and {Testi}, L. and {Natta}, A. and {Neri}, R. and {Cabrit}, S. and {Herczeg}, G.~J.},
        title = "{Dust properties of protoplanetary disks in the Taurus-Auriga star forming region from millimeter wavelengths}",
      journal = {\aap},
         year = 2010,
        month = mar,
       volume = {512},
          eid = {A15},
        pages = {A15},
          doi = {10.1051/0004-6361/200913403},
archivePrefix = {arXiv},
       eprint = {0912.3356},
 primaryClass = {astro-ph.EP},
       adsurl = {https://ui.adsabs.harvard.edu/abs/2010A&A...512A..15R}
}

@ARTICLE{2018ApJ...869L..45B,
       author = {{Birnstiel}, Tilman and {Dullemond}, Cornelis P. and {Zhu}, Zhaohuan and {Andrews}, Sean M. and {Bai}, Xue-Ning and {Wilner}, David J. and {Carpenter}, John M. and {Huang}, Jane and {Isella}, Andrea and {Benisty}, Myriam and {P{\'e}rez}, Laura M. and {Zhang}, Shangjia},
        title = "{The Disk Substructures at High Angular Resolution Project (DSHARP). V. Interpreting ALMA Maps of Protoplanetary Disks in Terms of a Dust Model}",
      journal = {\apjl},
         year = 2018,
        month = dec,
       volume = {869},
       number = {2},
          eid = {L45},
        pages = {L45},
          doi = {10.3847/2041-8213/aaf743},
archivePrefix = {arXiv},
       eprint = {1812.04043},
 primaryClass = {astro-ph.SR},
       adsurl = {https://ui.adsabs.harvard.edu/abs/2018ApJ...869L..45B}
}

@ARTICLE{2018ApJ...869L..46D,
       author = {{Dullemond}, Cornelis P. and {Birnstiel}, Tilman and {Huang}, Jane and {Kurtovic}, Nicol{\'a}s T. and {Andrews}, Sean M. and {Guzm{\'a}n}, Viviana V. and {P{\'e}rez}, Laura M. and {Isella}, Andrea and {Zhu}, Zhaohuan and {Benisty}, Myriam and {Wilner}, David J. and {Bai}, Xue-Ning and {Carpenter}, John M. and {Zhang}, Shangjia and {Ricci}, Luca},
        title = "{The Disk Substructures at High Angular Resolution Project (DSHARP). VI. Dust Trapping in Thin-ringed Protoplanetary Disks}",
      journal = {\apjl},
         year = 2018,
        month = dec,
       volume = {869},
       number = {2},
          eid = {L46},
        pages = {L46},
          doi = {10.3847/2041-8213/aaf742},
archivePrefix = {arXiv},
       eprint = {1812.04044},
 primaryClass = {astro-ph.EP},
       adsurl = {https://ui.adsabs.harvard.edu/abs/2018ApJ...869L..46D}
}

@ARTICLE{1996MNRAS.282.1321Z,
       author = {{Zubko}, V.~G. and {Mennella}, V. and {Colangeli}, L. and {Bussoletti}, E.},
        title = "{Optical constants of cosmic carbon analogue grains - I. Simulation of clustering by a modified continuous distribution of ellipsoids}",
      journal = {\mnras},
         year = 1996,
        month = oct,
       volume = {282},
       number = {4},
        pages = {1321-1329},
          doi = {10.1093/mnras/282.4.1321},
       adsurl = {https://ui.adsabs.harvard.edu/abs/1996MNRAS.282.1321Z}
}

@ARTICLE{2001ApJ...553..321D,
       author = {{D'Alessio}, Paola and {Calvet}, Nuria and {Hartmann}, Lee},
        title = "{Accretion Disks around Young Objects. III. Grain Growth}",
      journal = {\apj},
         year = 2001,
        month = may,
       volume = {553},
       number = {1},
        pages = {321-334},
          doi = {10.1086/320655},
archivePrefix = {arXiv},
       eprint = {astro-ph/0101443},
 primaryClass = {astro-ph},
       adsurl = {https://ui.adsabs.harvard.edu/abs/2001ApJ...553..321D}
}

@ARTICLE{1998ApJ...500..411D,
       author = {{D'Alessio}, Paola and {Cant{\"o}}, Jorge and {Calvet}, Nuria and {Lizano}, Susana},
        title = "{Accretion Disks around Young Objects. I. The Detailed Vertical Structure}",
      journal = {\apj},
         year = 1998,
        month = jun,
       volume = {500},
       number = {1},
        pages = {411-427},
          doi = {10.1086/305702},
archivePrefix = {arXiv},
       eprint = {astro-ph/9806060},
 primaryClass = {astro-ph},
       adsurl = {https://ui.adsabs.harvard.edu/abs/1998ApJ...500..411D}
}

@ARTICLE{2016ApJ...817...17G,
       author = {{Greene}, Thomas P. and {Line}, Michael R. and {Montero}, Cezar and {Fortney}, Jonathan J. and {Lustig-Yaeger}, Jacob and {Luther}, Kyle},
        title = "{Characterizing Transiting Exoplanet Atmospheres with JWST}",
      journal = {\apj},
         year = 2016,
        month = jan,
       volume = {817},
       number = {1},
          eid = {17},
        pages = {17},
          doi = {10.3847/0004-637X/817/1/17},
archivePrefix = {arXiv},
       eprint = {1511.05528},
 primaryClass = {astro-ph.EP},
       adsurl = {https://ui.adsabs.harvard.edu/abs/2016ApJ...817...17G}
}

@ARTICLE{2019ARA&A..57..617M,
       author = {{Madhusudhan}, Nikku},
        title = "{Exoplanetary Atmospheres: Key Insights, Challenges, and Prospects}",
      journal = {\araa},
         year = 2019,
        month = aug,
       volume = {57},
        pages = {617-663},
          doi = {10.1146/annurev-astro-081817-051846},
archivePrefix = {arXiv},
       eprint = {1904.03190},
 primaryClass = {astro-ph.EP},
       adsurl = {https://ui.adsabs.harvard.edu/abs/2019ARA&A..57..617M}
}

@ARTICLE{2013ApJ...763...25M,
       author = {{Moses}, J.~I. and {Madhusudhan}, N. and {Visscher}, C. and {Freedman}, R.~S.},
        title = "{Chemical Consequences of the C/O Ratio on Hot Jupiters: Examples from WASP-12b, CoRoT-2b, XO-1b, and HD 189733b}",
      journal = {\apj},
         year = 2013,
        month = jan,
       volume = {763},
       number = {1},
          eid = {25},
        pages = {25},
          doi = {10.1088/0004-637X/763/1/25},
archivePrefix = {arXiv},
       eprint = {1211.2996},
 primaryClass = {astro-ph.EP},
       adsurl = {https://ui.adsabs.harvard.edu/abs/2013ApJ...763...25M}
}

@ARTICLE{2012ApJ...758...36M,
       author = {{Madhusudhan}, Nikku},
        title = "{C/O Ratio as a Dimension for Characterizing Exoplanetary Atmospheres}",
      journal = {\apj},
         year = 2012,
        month = oct,
       volume = {758},
       number = {1},
          eid = {36},
        pages = {36},
          doi = {10.1088/0004-637X/758/1/36},
archivePrefix = {arXiv},
       eprint = {1209.2412},
 primaryClass = {astro-ph.EP},
       adsurl = {https://ui.adsabs.harvard.edu/abs/2012ApJ...758...36M}
}

@ARTICLE{2024ApJ...969L..21B,
       author = {{Bergin}, Edwin A. and {Booth}, Richard A. and {Colmenares}, Maria Jose and {Ilee}, John D.},
        title = "{C/O Ratios and the Formation of Wide-separation Exoplanets}",
      journal = {\apjl},
         year = 2024,
        month = jul,
       volume = {969},
       number = {1},
          eid = {L21},
        pages = {L21},
          doi = {10.3847/2041-8213/ad5839},
archivePrefix = {arXiv},
       eprint = {2406.12037},
 primaryClass = {astro-ph.EP},
       adsurl = {https://ui.adsabs.harvard.edu/abs/2024ApJ...969L..21B}
}

@ARTICLE{2021PSJ.....2...25W,
       author = {{Woodward}, Charles E. and {Wooden}, Diane H. and {Harker}, David E. and {Kelley}, Michael S.~P. and {Russell}, Ray W. and {Kim}, Daryl L.},
        title = "{The Coma Dust of Comet C/2013 US$_{10}$ (Catalina): A Window into Carbon in the Solar System}",
      journal = {\psj},
         year = 2021,
        month = feb,
       volume = {2},
       number = {1},
          eid = {25},
        pages = {25},
          doi = {10.3847/PSJ/abca3e},
archivePrefix = {arXiv},
       eprint = {2011.06943},
 primaryClass = {astro-ph.EP},
       adsurl = {https://ui.adsabs.harvard.edu/abs/2021PSJ.....2...25W}
}

@ARTICLE{2021A&A...653A.141A,
       author = {{Asplund}, M. and {Amarsi}, A.~M. and {Grevesse}, N.},
        title = "{The chemical make-up of the Sun: A 2020 vision}",
      journal = {\aap},
         year = 2021,
        month = sep,
       volume = {653},
          eid = {A141},
        pages = {A141},
          doi = {10.1051/0004-6361/202140445},
archivePrefix = {arXiv},
       eprint = {2105.01661},
 primaryClass = {astro-ph.SR},
       adsurl = {https://ui.adsabs.harvard.edu/abs/2021A&A...653A.141A}
}

@ARTICLE{2021MNRAS.508.2236B,
       author = {{Bergemann}, Maria and {Hoppe}, Richard and {Semenova}, Ekaterina and {Carlsson}, Mats and {Yakovleva}, Svetlana A. and {Voronov}, Yaroslav V. and {Bautista}, Manuel and {Nemer}, Ahmad and {Belyaev}, Andrey K. and {Leenaarts}, Jorrit and {Mashonkina}, Lyudmila and {Reiners}, Ansgar and {Ellwarth}, Monika},
        title = "{Solar oxygen abundance}",
      journal = {\mnras},
         year = 2021,
        month = dec,
       volume = {508},
       number = {2},
        pages = {2236-2253},
          doi = {10.1093/mnras/stab2160},
archivePrefix = {arXiv},
       eprint = {2109.01143},
 primaryClass = {astro-ph.SR},
       adsurl = {https://ui.adsabs.harvard.edu/abs/2021MNRAS.508.2236B}
}

@ARTICLE{2024A&A...688A.193D,
       author = {{da Silva}, R. and {Danielski}, C. and {Delgado Mena}, E. and {Magrini}, L. and {Turrini}, D. and {Biazzo}, K. and {Tsantaki}, M. and {Rainer}, M. and {Helminiak}, K.~G. and {Benatti}, S. and {Adibekyan}, V. and {Sanna}, N. and {Sousa}, S. and {Casali}, G. and {Van der Swaelmen}, M.},
        title = "{Ariel stellar characterisation. II. Chemical abundances of carbon, nitrogen, and oxygen for 181 planet-host FGK dwarf stars}",
      journal = {\aap},
         year = 2024,
        month = aug,
       volume = {688},
          eid = {A193},
        pages = {A193},
          doi = {10.1051/0004-6361/202450604},
archivePrefix = {arXiv},
       eprint = {2406.12393},
 primaryClass = {astro-ph.SR},
       adsurl = {https://ui.adsabs.harvard.edu/abs/2024A&A...688A.193D}
}

@ARTICLE{2012ApJ...747L..27F,
       author = {{Fortney}, Jonathan J.},
        title = "{On the Carbon-to-oxygen Ratio Measurement in nearby Sun-like Stars: Implications for Planet Formation and the Determination of Stellar Abundances}",
      journal = {\apjl},
         year = 2012,
        month = mar,
       volume = {747},
       number = {2},
          eid = {L27},
        pages = {L27},
          doi = {10.1088/2041-8205/747/2/L27},
archivePrefix = {arXiv},
       eprint = {1201.1504},
 primaryClass = {astro-ph.SR},
       adsurl = {https://ui.adsabs.harvard.edu/abs/2012ApJ...747L..27F}
}

@ARTICLE{Nissen2013,
       author = {{Nissen}, P.~E.},
        title = "{The carbon-to-oxygen ratio in stars with planets}",
      journal = {\aap},
         year = 2013,
        month = apr,
       volume = {552},
          eid = {A73},
        pages = {A73},
          doi = {10.1051/0004-6361/201321234},
archivePrefix = {arXiv},
       eprint = {1303.1726},
 primaryClass = {astro-ph.SR},
       adsurl = {https://ui.adsabs.harvard.edu/abs/2013A&A...552A..73N}
}

@ARTICLE{Hypatia,
       author = {{Hinkel}, Natalie R. and {Timmes}, F.~X. and {Young}, Patrick A. and {Pagano}, Michael D. and {Turnbull}, Margaret C.},
        title = "{Stellar Abundances in the Solar Neighborhood: The Hypatia Catalog}",
      journal = {\aj},
         year = 2014,
        month = sep,
       volume = {148},
       number = {3},
          eid = {54},
        pages = {54},
          doi = {10.1088/0004-6256/148/3/54},
archivePrefix = {arXiv},
       eprint = {1405.6719},
 primaryClass = {astro-ph.SR},
       adsurl = {https://ui.adsabs.harvard.edu/abs/2014AJ....148...54H}
}

@article{NERI2020,
title = {A carbonaceous chondrite and cometary origin for icy moons of Jupiter and Saturn},
journal = {Earth and Planetary Science Letters},
volume = {530},
pages = {115920},
year = {2020},
issn = {0012-821X},
doi = {https://doi.org/10.1016/j.epsl.2019.115920},
url = {https://www.sciencedirect.com/science/article/pii/S0012821X19306120},
author = {Adrien Néri and François Guyot and Bruno Reynard and Christophe Sotin}
}

@article{Gao2010methane,
    author = {Gao, Guoying and Oganov, Artem R. and Ma, Yanming and Wang, Hui and Li, Peifang and Li, Yinwei and Iitaka, Toshiaki and Zou, Guangtian},
    title = {Dissociation of methane under high pressure},
    journal = {The Journal of Chemical Physics},
    volume = {133},
    number = {14},
    pages = {144508},
    year = {2010},
    month = {10},
    issn = {0021-9606},
    doi = {10.1063/1.3488102},
}

@ARTICLE{Vinet1989,
       author = {{Vinet}, Pascal and {Rose}, James H. and {Ferrante}, John and {Smith}, John R.},
        title = "{Universal features of the equation of state of solids}",
      journal = {Journal of Physics Condensed Matter},
         year = 1989,
        month = mar,
       volume = {1},
       number = {11},
        pages = {1941-1963},
          doi = {10.1088/0953-8984/1/11/002},
       adsurl = {https://ui.adsabs.harvard.edu/abs/1989JPCM....1.1941V}
}

@article{Tkacz2002,
author = {Tkacz, Marek and Litwiniuk, A},
year = {2002},
month = {01},
pages = {89-92},
title = {Useful equations of state of hydrogen and deuterium},
volume = {330},
journal = {Journal of Alloys and Compounds - J ALLOYS COMPOUNDS},
doi = {10.1016/S0925-8388(01)01488-8}
}

@article{Davydov2004,
  title={Conversion of polycyclic aromatic hydrocarbons to graphite and diamond at high pressures},
  author={{Davydov}, V.~A. and {Rakhmanina}, A.~V. and {Agafonov}, V. and {Narymbetov}, B. and {Boudou}, J.-P. and {Szwarc}, H.},
  journal={Carbon},
  volume={42},
  number={2},
  pages={261--269},
  year={2004},
  publisher={Elsevier}
}

@ARTICLE{Mangan2017,
       author = {{Mangan}, T.~P. and {Salzmann}, C.~G. and {Plane}, J.~M.~C. and {Murray}, B.~J.},
        title = "{CO$_{2}$ ice structure and density under Martian atmospheric conditions}",
      journal = {\icarus},
         year = 2017,
        month = sep,
       volume = {294},
        pages = {201-208},
          doi = {10.1016/j.icarus.2017.03.012},
       adsurl = {https://ui.adsabs.harvard.edu/abs/2017Icar..294..201M}
}

@ARTICLE{Owen2013,
       author = {{Owen}, James E. and {Wu}, Yanqin},
        title = "{Kepler Planets: A Tale of Evaporation}",
      journal = {\apj},
         year = 2013,
        month = oct,
       volume = {775},
       number = {2},
          eid = {105},
        pages = {105},
          doi = {10.1088/0004-637X/775/2/105},
archivePrefix = {arXiv},
       eprint = {1303.3899},
 primaryClass = {astro-ph.EP},
       adsurl = {https://ui.adsabs.harvard.edu/abs/2013ApJ...775..105O}
}

@INPROCEEDINGS{Lichtenberg2023,
       author = {{Lichtenberg}, Tim and {Schaefer}, Laura K. and {Nakajima}, Miki and {Fischer}, Rebecca A.},
        title = "{Geophysical Evolution During Rocky Planet Formation}",
    booktitle = {Protostars and Planets VII},
         year = 2023,
       editor = {{Inutsuka}, S. and {Aikawa}, Y. and {Muto}, T. and {Tomida}, K. and {Tamura}, M.},
       series = {Astronomical Society of the Pacific Conference Series},
       volume = {534},
        month = jul,
        pages = {907},
          doi = {10.48550/arXiv.2203.10023},
archivePrefix = {arXiv},
       eprint = {2203.10023},
 primaryClass = {astro-ph.EP},
       adsurl = {https://ui.adsabs.harvard.edu/abs/2023ASPC..534..907L}
}

@ARTICLE{Zeng2019,
       author = {{Zeng}, Li and {Jacobsen}, Stein B. and {Sasselov}, Dimitar D. and {Petaev}, Michail I. and {Vanderburg}, Andrew and {Lopez-Morales}, Mercedes and {Perez-Mercader}, Juan and {Mattsson}, Thomas R. and {Li}, Gongjie and {Heising}, Matthew Z. and {Bonomo}, Aldo S. and {Damasso}, Mario and {Berger}, Travis A. and {Cao}, Hao and {Levi}, Amit and {Wordsworth}, Robin D.},
        title = "{Growth model interpretation of planet size distribution}",
      journal = {Proceedings of the National Academy of Science},
         year = 2019,
        month = may,
       volume = {116},
       number = {20},
        pages = {9723-9728},
          doi = {10.1073/pnas.1812905116},
archivePrefix = {arXiv},
       eprint = {1906.04253},
 primaryClass = {astro-ph.EP},
       adsurl = {https://ui.adsabs.harvard.edu/abs/2019PNAS..116.9723Z}
}

@ARTICLE{Bradley2009,
       author = {{Bradley}, D.~K. and {Eggert}, J.~H. and {Smith}, R.~F. and {Prisbrey}, S.~T. and {Hicks}, D.~G. and {Braun}, D.~G. and {Biener}, J. and {Hamza}, A.~V. and {Rudd}, R.~E. and {Collins}, G.~W.},
        title = "{Diamond at 800GPa}",
      journal = {\prl},
         year = 2009,
        month = feb,
       volume = {102},
       number = {7},
          eid = {075503},
        pages = {075503},
          doi = {10.1103/PhysRevLett.102.075503},
       adsurl = {https://ui.adsabs.harvard.edu/abs/2009PhRvL.102g5503B}
}

@ARTICLE{Stixrude2021,
       author = {{Stixrude}, Lars and {Baroni}, Stefano and {Grasselli}, Federico},
        title = "{Thermal and Tidal Evolution of Uranus with a Growing Frozen Core}",
      journal = {\psj},
         year = 2021,
        month = dec,
       volume = {2},
       number = {6},
          eid = {222},
        pages = {222},
          doi = {10.3847/PSJ/ac2a47},
       adsurl = {https://ui.adsabs.harvard.edu/abs/2021PSJ.....2..222S}
}

@ARTICLE{Fei2021,
       author = {{Fei}, Yingwei and {Seagle}, Christopher T. and {Townsend}, Joshua P. and {McCoy}, Chad A. and {Boujibar}, Asmaa and {Driscoll}, Peter and {Shulenburger}, Luke and {Furnish}, Michael D.},
        title = "{Melting and density of MgSiO$_{3}$ determined by shock compression of bridgmanite to 1254GPa}",
      journal = {Nature Communications},
         year = 2021,
        month = jan,
       volume = {12},
          eid = {876},
        pages = {876},
          doi = {10.1038/s41467-021-21170-y},
       adsurl = {https://ui.adsabs.harvard.edu/abs/2021NatCo..12..876F}
}

@ARTICLE{2022arXiv220316065S,
       author = {{Swift}, Damian C. and {Kritcher}, Andrea L. and {Lazicki}, Amy and {Hawreliak}, James A. and {Doeppner}, Tilo and {Whitley}, Heather D. and {Nilsen}, Joseph and {Bachmann}, Benjamin and {MacDonald}, Michael and {Maddox}, Brian and {Kostinski}, Natalie and {Collins}, Gilbert W. and {Glenzer}, Siegfried and {Rothman}, Stephen D. and {Kraus}, Dominik and {Falcone}, Roger W.},
        title = "{Shock Hugoniot of diamond from 3 to 80 TPa}",
      journal = {arXiv e-prints},
         year = 2022,
        month = mar,
          eid = {arXiv:2203.16065},
        pages = {arXiv:2203.16065},
          doi = {10.48550/arXiv.2203.16065},
archivePrefix = {arXiv},
       eprint = {2203.16065},
 primaryClass = {physics.plasm-ph},
       adsurl = {https://ui.adsabs.harvard.edu/abs/2022arXiv220316065S}
}

@ARTICLE{Smith2018_fe_eos,
       author = {{Smith}, Raymond F. and {Fratanduono}, Dayne E. and {Braun}, David G. and {Duffy}, Thomas S. and {Wicks}, June K. and {Celliers}, Peter M. and {Ali}, Suzanne J. and {Fernandez-Pa{\~n}ella}, Amalia and {Kraus}, Richard G. and {Swift}, Damian C. and {Collins}, Gilbert W. and {Eggert}, Jon H.},
        title = "{Equation of state of iron under core conditions of large rocky exoplanets}",
      journal = {Nature Astronomy},
         year = 2018,
        month = apr,
       volume = {2},
        pages = {452-458},
          doi = {10.1038/s41550-018-0437-9},
       adsurl = {https://ui.adsabs.harvard.edu/abs/2018NatAs...2..452S}
}

@ARTICLE{Sun2009,
       author = {{Sun}, Liling and {Yi}, Wei and {Wang}, Lin and {Shu}, Jinfu and {Sinogeikin}, Stas and {Meng}, Yue and {Shen}, Guoyin and {Bai}, Ligang and {Li}, Yanchuan and {Liu}, Jing and {Mao}, Ho-kwang and {Mao}, Wendy L.},
        title = "{X-ray diffraction studies and equation of state of methane at 202 GPa}",
      journal = {Chemical Physics Letters},
         year = 2009,
        month = apr,
       volume = {473},
       number = {1},
        pages = {72-74},
          doi = {10.1016/j.cplett.2009.03.072},
       adsurl = {https://ui.adsabs.harvard.edu/abs/2009CPL...473...72S}
}

@ARTICLE{blanchard22,
       author = {{Blanchard}, I. and {Rubie}, D.~C. and {Jennings}, E.~S. and {Franchi}, I.~A. and {Zhao}, X. and {Petitgirard}, S. and {Miyajima}, N. and {Jacobson}, S.~A. and {Morbidelli}, A.},
        title = "{The metal-silicate partitioning of carbon during Earth's accretion and its distribution in the early solar system}",
      journal = {Earth and Planetary Science Letters},
         year = 2022,
        month = feb,
       volume = {580},
          eid = {117374},
        pages = {117374},
          doi = {10.1016/j.epsl.2022.117374},
archivePrefix = {arXiv},
       eprint = {2202.06809},
 primaryClass = {astro-ph.EP},
       adsurl = {https://ui.adsabs.harvard.edu/abs/2022E&PSL.58017374B}
}

@ARTICLE{seo24,
       author = {{Seo}, Chanoul and {Ito}, Yuichi and {Fujii}, Yuka},
        title = "{Role of magma oceans in controlling carbon and oxygen of sub-Neptune atmospheres}",
      journal = {arXiv e-prints},
         year = 2024,
        month = aug,
          eid = {arXiv:2408.17056},
        pages = {arXiv:2408.17056},
          doi = {10.48550/arXiv.2408.17056},
archivePrefix = {arXiv},
       eprint = {2408.17056},
 primaryClass = {astro-ph.EP},
       adsurl = {https://ui.adsabs.harvard.edu/abs/2024arXiv240817056S}
}

@ARTICLE{Binkert2023,
       author = {{Binkert}, Fabian and {Birnstiel}, Til},
        title = "{Carbon depletion in the early Solar system}",
      journal = {\mnras},
         year = 2023,
        month = apr,
       volume = {520},
       number = {2},
        pages = {2055-2080},
          doi = {10.1093/mnras/stad182},
archivePrefix = {arXiv},
       eprint = {2301.05706},
 primaryClass = {astro-ph.EP},
       adsurl = {https://ui.adsabs.harvard.edu/abs/2023MNRAS.520.2055B}
}

@ARTICLE{Seager2007,
       author = {{Seager}, S. and {Kuchner}, M. and {Hier-Majumder}, C.~A. and {Militzer}, B.},
        title = "{Mass-Radius Relationships for Solid Exoplanets}",
      journal = {\apj},
         year = 2007,
        month = nov,
       volume = {669},
       number = {2},
        pages = {1279-1297},
          doi = {10.1086/521346},
archivePrefix = {arXiv},
       eprint = {0707.2895},
 primaryClass = {astro-ph},
       adsurl = {https://ui.adsabs.harvard.edu/abs/2007ApJ...669.1279S}
}

@ARTICLE{wogan24,
       author = {{Wogan}, Nicholas F. and {Batalha}, Natasha E. and {Zahnle}, Kevin J. and {Krissansen-Totton}, Joshua and {Tsai}, Shang-Min and {Hu}, Renyu},
        title = "{JWST Observations of K2-18b Can Be Explained by a Gas-rich Mini-Neptune with No Habitable Surface}",
      journal = {\apjl},
         year = 2024,
        month = mar,
       volume = {963},
       number = {1},
          eid = {L7},
        pages = {L7},
          doi = {10.3847/2041-8213/ad2616},
archivePrefix = {arXiv},
       eprint = {2401.11082},
 primaryClass = {astro-ph.EP},
       adsurl = {https://ui.adsabs.harvard.edu/abs/2024ApJ...963L...7W}
}

@ARTICLE{rigby24,
       author = {{Rigby}, Frances E. and {Pica-Ciamarra}, Lorenzo and {Holmberg}, M{\r{a}}ns and {Madhusudhan}, Nikku and {Constantinou}, Savvas and {Schaefer}, Laura and {Deng}, Jie and {Lee}, Kanani K.~M. and {Moses}, Julianne I.},
        title = "{Towards a self-consistent evaluation of gas dwarf scenarios for temperate sub-Neptunes}",
      journal = {arXiv e-prints},
         year = 2024,
        month = sep,
          eid = {arXiv:2409.03683},
        pages = {arXiv:2409.03683},
          doi = {10.48550/arXiv.2409.03683},
archivePrefix = {arXiv},
       eprint = {2409.03683},
 primaryClass = {astro-ph.EP},
       adsurl = {https://ui.adsabs.harvard.edu/abs/2024arXiv240903683R}
}

@ARTICLE{nixon24,
       author = {{Nixon}, Matthew C. and {Piette}, Anjali A.~A. and {Kempton}, Eliza M. -R. and {Gao}, Peter and {Bean}, Jacob L. and {Steinrueck}, Maria E. and {Mahajan}, Alexandra S. and {Eastman}, Jason D. and {Zhang}, Michael and {Rogers}, Leslie A.},
        title = "{New Insights into the Internal Structure of GJ 1214 b Informed by JWST}",
      journal = {\apjl},
         year = 2024,
        month = aug,
       volume = {970},
       number = {2},
          eid = {L28},
        pages = {L28},
          doi = {10.3847/2041-8213/ad615b},
archivePrefix = {arXiv},
       eprint = {2407.12079},
 primaryClass = {astro-ph.EP},
       adsurl = {https://ui.adsabs.harvard.edu/abs/2024ApJ...970L..28N}
}

@ARTICLE{Neish2024,
       author = {{Neish}, Catherine and {Malaska}, Michael J. and {Sotin}, Christophe and {Lopes}, Rosaly M.~C. and {Nixon}, Conor A. and {Affholder}, Antonin and {Chatain}, Audrey and {Cockell}, Charles and {Farnsworth}, Kendra K. and {Higgins}, Peter M. and {Miller}, Kelly E. and {Soderlund}, Krista M.},
        title = "{Organic Input to Titan's Subsurface Ocean Through Impact Cratering}",
      journal = {Astrobiology},
         year = 2024,
        month = feb,
       volume = {24},
       number = {2},
        pages = {177-189},
          doi = {10.1089/ast.2023.0055},
       adsurl = {https://ui.adsabs.harvard.edu/abs/2024AsBio..24..177N}
}

@ARTICLE{Cleaves2014,
       author = {{Cleaves}, H. James and {Neish}, Catherine and {Callahan}, Michael P. and {Parker}, Eric and {Fern{\'a}ndez}, Facundo M. and {Dworkin}, Jason P.},
        title = "{Amino acids generated from hydrated Titan tholins: Comparison with Miller-Urey electric discharge products}",
      journal = {\icarus},
         year = 2014,
        month = jul,
       volume = {237},
        pages = {182-189},
          doi = {10.1016/j.icarus.2014.04.042},
       adsurl = {https://ui.adsabs.harvard.edu/abs/2014Icar..237..182C}
}

@ARTICLE{holmberg24,
       author = {{Holmberg}, M{\r{a}}ns and {Madhusudhan}, Nikku},
        title = "{Possible Hycean conditions in the sub-Neptune TOI-270 d}",
      journal = {\aap},
         year = 2024,
        month = mar,
       volume = {683},
          eid = {L2},
        pages = {L2},
          doi = {10.1051/0004-6361/202348238},
archivePrefix = {arXiv},
       eprint = {2403.03244},
 primaryClass = {astro-ph.EP},
       adsurl = {https://ui.adsabs.harvard.edu/abs/2024A&A...683L...2H}
}

@ARTICLE{morley13,
       author = {{Morley}, Caroline V. and {Fortney}, Jonathan J. and {Kempton}, Eliza M. -R. and {Marley}, Mark S. and {Visscher}, Channon and {Zahnle}, Kevin},
        title = "{Quantitatively Assessing the Role of Clouds in the Transmission Spectrum of GJ 1214b}",
      journal = {\apj},
         year = 2013,
        month = sep,
       volume = {775},
       number = {1},
          eid = {33},
        pages = {33},
          doi = {10.1088/0004-637X/775/1/33},
archivePrefix = {arXiv},
       eprint = {1305.4124},
 primaryClass = {astro-ph.EP},
       adsurl = {https://ui.adsabs.harvard.edu/abs/2013ApJ...775...33M}
}

@ARTICLE{millerricci12,
       author = {{Miller-Ricci Kempton}, Eliza and {Zahnle}, Kevin and {Fortney}, Jonathan J.},
        title = "{The Atmospheric Chemistry of GJ 1214b: Photochemistry and Clouds}",
      journal = {\apj},
         year = 2012,
        month = jan,
       volume = {745},
       number = {1},
          eid = {3},
        pages = {3},
          doi = {10.1088/0004-637X/745/1/3},
archivePrefix = {arXiv},
       eprint = {1104.5477},
 primaryClass = {astro-ph.EP},
       adsurl = {https://ui.adsabs.harvard.edu/abs/2012ApJ...745....3M}
}

@ARTICLE{2012PhRvL.108i1102K,
       author = {{Knudson}, M.~D. and {Desjarlais}, M.~P. and {Lemke}, R.~W. and {Mattsson}, T.~R. and {French}, M. and {Nettelmann}, N. and {Redmer}, R.},
        title = "{Probing the Interiors of the Ice Giants: Shock Compression of Water to 700 GPa and 3.8g/cm$^{3}$}",
      journal = {\prl},
         year = 2012,
        month = mar,
       volume = {108},
       number = {9},
          eid = {091102},
        pages = {091102},
          doi = {10.1103/PhysRevLett.108.091102},
archivePrefix = {arXiv},
       eprint = {1201.2622},
 primaryClass = {astro-ph.EP},
       adsurl = {https://ui.adsabs.harvard.edu/abs/2012PhRvL.108i1102K}
}

@ARTICLE{benneke24,
       author = {{Benneke}, Bj{\"o}rn and {Roy}, Pierre-Alexis and {Coulombe}, Louis-Philippe and {Radica}, Michael and {Piaulet}, Caroline and {Ahrer}, Eva-Maria and {Pierrehumbert}, Raymond and {Krissansen-Totton}, Joshua and {Schlichting}, Hilke E. and {Hu}, Renyu and {Yang}, Jeehyun and {Christie}, Duncan and {Thorngren}, Daniel and {Young}, Edward D. and {Pelletier}, Stefan and {Knutson}, Heather A. and {Miguel}, Yamila and {Evans-Soma}, Thomas M. and {Dorn}, Caroline and {Gagnebin}, Anna and {Fortney}, Jonathan J. and {Komacek}, Thaddeus and {MacDonald}, Ryan and {Raul}, Eshan and {Cloutier}, Ryan and {Acuna}, Lorena and {Lafreni{\`e}re}, David and {Cadieux}, Charles and {Doyon}, Ren{\'e} and {Welbanks}, Luis and {Allart}, Romain},
        title = "{JWST Reveals CH$_4$, CO$_2$, and H$_2$O in a Metal-rich Miscible Atmosphere on a Two-Earth-Radius Exoplanet}",
      journal = {arXiv e-prints},
         year = 2024,
        month = mar,
          eid = {arXiv:2403.03325},
        pages = {arXiv:2403.03325},
          doi = {10.48550/arXiv.2403.03325},
archivePrefix = {arXiv},
       eprint = {2403.03325},
 primaryClass = {astro-ph.EP},
       adsurl = {https://ui.adsabs.harvard.edu/abs/2024arXiv240303325B}
}

@ARTICLE{madhu23,
       author = {{Madhusudhan}, Nikku and {Sarkar}, Subhajit and {Constantinou}, Savvas and {Holmberg}, M{\r{a}}ns and {Piette}, Anjali A.~A. and {Moses}, Julianne I.},
        title = "{Carbon-bearing Molecules in a Possible Hycean Atmosphere}",
      journal = {\apjl},
         year = 2023,
        month = oct,
       volume = {956},
       number = {1},
          eid = {L13},
        pages = {L13},
          doi = {10.3847/2041-8213/acf577},
archivePrefix = {arXiv},
       eprint = {2309.05566},
 primaryClass = {astro-ph.EP},
       adsurl = {https://ui.adsabs.harvard.edu/abs/2023ApJ...956L..13M}
}

@ARTICLE{madhu21,
       author = {{Madhusudhan}, Nikku and {Piette}, Anjali A.~A. and {Constantinou}, Savvas},
        title = "{Habitability and Biosignatures of Hycean Worlds}",
      journal = {\apj},
         year = 2021,
        month = sep,
       volume = {918},
       number = {1},
          eid = {1},
        pages = {1},
          doi = {10.3847/1538-4357/abfd9c},
archivePrefix = {arXiv},
       eprint = {2108.10888},
 primaryClass = {astro-ph.EP},
       adsurl = {https://ui.adsabs.harvard.edu/abs/2021ApJ...918....1M}
}

@ARTICLE{rogers10,
       author = {{Rogers}, L.~A. and {Seager}, S.},
        title = "{Three Possible Origins for the Gas Layer on GJ 1214b}",
      journal = {\apj},
         year = 2010,
        month = jun,
       volume = {716},
       number = {2},
        pages = {1208-1216},
          doi = {10.1088/0004-637X/716/2/1208},
archivePrefix = {arXiv},
       eprint = {0912.3243},
 primaryClass = {astro-ph.EP},
       adsurl = {https://ui.adsabs.harvard.edu/abs/2010ApJ...716.1208R}
}

@ARTICLE{Loubeyre1996,
       author = {{Loubeyre}, P. and {Letoullec}, R. and {Hausermann}, D. and {Hanfland}, M. and {Hemley}, R.~J. and {Mao}, H.~K. and {Finger}, L.~W.},
        title = "{X-ray diffraction and equation of state of hydrogen at megabar pressures}",
      journal = {\nat},
         year = 1996,
        month = oct,
       volume = {383},
       number = {6602},
        pages = {702-704},
          doi = {10.1038/383702a0},
       adsurl = {https://ui.adsabs.harvard.edu/abs/1996Natur.383..702L}
}

@ARTICLE{owen17,
       author = {{Owen}, James E. and {Wu}, Yanqin},
        title = "{The Evaporation Valley in the Kepler Planets}",
      journal = {\apj},
         year = 2017,
        month = sep,
       volume = {847},
       number = {1},
          eid = {29},
        pages = {29},
          doi = {10.3847/1538-4357/aa890a},
archivePrefix = {arXiv},
       eprint = {1705.10810},
 primaryClass = {astro-ph.EP},
       adsurl = {https://ui.adsabs.harvard.edu/abs/2017ApJ...847...29O}
}

@ARTICLE{fulton17,
       author = {{Fulton}, Benjamin J. and {Petigura}, Erik A. and {Howard}, Andrew W. and {Isaacson}, Howard and {Marcy}, Geoffrey W. and {Cargile}, Phillip A. and {Hebb}, Leslie and {Weiss}, Lauren M. and {Johnson}, John Asher and {Morton}, Timothy D. and {Sinukoff}, Evan and {Crossfield}, Ian J.~M. and {Hirsch}, Lea A.},
        title = "{The California-Kepler Survey. III. A Gap in the Radius Distribution of Small Planets}",
      journal = {\aj},
         year = 2017,
        month = sep,
       volume = {154},
       number = {3},
          eid = {109},
        pages = {109},
          doi = {10.3847/1538-3881/aa80eb},
archivePrefix = {arXiv},
       eprint = {1703.10375},
 primaryClass = {astro-ph.EP},
       adsurl = {https://ui.adsabs.harvard.edu/abs/2017AJ....154..109F}
}

@ARTICLE{Charbonneau2009,
       author = {{Charbonneau}, David and {Berta}, Zachory K. and {Irwin}, Jonathan and {Burke}, Christopher J. and {Nutzman}, Philip and {Buchhave}, Lars A. and {Lovis}, Christophe and {Bonfils}, Xavier and {Latham}, David W. and {Udry}, St{\'e}phane and {Murray-Clay}, Ruth A. and {Holman}, Matthew J. and {Falco}, Emilio E. and {Winn}, Joshua N. and {Queloz}, Didier and {Pepe}, Francesco and {Mayor}, Michel and {Delfosse}, Xavier and {Forveille}, Thierry},
        title = "{A super-Earth transiting a nearby low-mass star}",
      journal = {\nat},
         year = 2009,
        month = dec,
       volume = {462},
       number = {7275},
        pages = {891-894},
          doi = {10.1038/nature08679},
archivePrefix = {arXiv},
       eprint = {0912.3229},
 primaryClass = {astro-ph.EP},
       adsurl = {https://ui.adsabs.harvard.edu/abs/2009Natur.462..891C}
}

@ARTICLE{Bardyn2017,
       author = {{Bardyn}, Ana{\"\i}s and {Baklouti}, Donia and {Cottin}, Herv{\'e} and {Fray}, Nicolas and {Briois}, Christelle and {Paquette}, John and {Stenzel}, Oliver and {Engrand}, C{\'e}cile and {Fischer}, Henning and {Hornung}, Klaus and {Isnard}, Robin and {Langevin}, Yves and {Lehto}, Harry and {Le Roy}, L{\'e}na and {Ligier}, Nicolas and {Merouane}, Sihane and {Modica}, Paola and {Orthous-Daunay}, Fran{\c{c}}ois-R{\'e}gis and {Ryn{\"o}}, Jouni and {Schulz}, Rita and {Sil{\'e}n}, Johan and {Thirkell}, Laurent and {Varmuza}, Kurt and {Zaprudin}, Boris and {Kissel}, Jochen and {Hilchenbach}, Martin},
        title = "{Carbon-rich dust in comet 67P/Churyumov-Gerasimenko measured by COSIMA/Rosetta}",
      journal = {\mnras},
         year = 2017,
        month = jul,
       volume = {469},
        pages = {S712-S722},
          doi = {10.1093/mnras/stx2640},
       adsurl = {https://ui.adsabs.harvard.edu/abs/2017MNRAS.469S.712B}
}

@ARTICLE{Borucki2011,
       author = {{Borucki}, William J. and {Koch}, David G. and {Basri}, Gibor and {Batalha}, Natalie and {Brown}, Timothy M. and {Bryson}, Stephen T. and {Caldwell}, Douglas and {Christensen-Dalsgaard}, J{\o}rgen and {Cochran}, William D. and {DeVore}, Edna and {Dunham}, Edward W. and {Gautier}, Thomas N., III and {Geary}, John C. and {Gilliland}, Ronald and {Gould}, Alan and {Howell}, Steve B. and {Jenkins}, Jon M. and {Latham}, David W. and {Lissauer}, Jack J. and {Marcy}, Geoffrey W. and {Rowe}, Jason and {Sasselov}, Dimitar and {Boss}, Alan and {Charbonneau}, David and {Ciardi}, David and {Doyle}, Laurance and {Dupree}, Andrea K. and {Ford}, Eric B. and {Fortney}, Jonathan and {Holman}, Matthew J. and {Seager}, Sara and {Steffen}, Jason H. and {Tarter}, Jill and {Welsh}, William F. and {Allen}, Christopher and {Buchhave}, Lars A. and {Christiansen}, Jessie L. and {Clarke}, Bruce D. and {Das}, Santanu and {D{\'e}sert}, Jean-Michel and {Endl}, Michael and {Fabrycky}, Daniel and {Fressin}, Francois and {Haas}, Michael and {Horch}, Elliott and {Howard}, Andrew and {Isaacson}, Howard and {Kjeldsen}, Hans and {Kolodziejczak}, Jeffery and {Kulesa}, Craig and {Li}, Jie and {Lucas}, Philip W. and {Machalek}, Pavel and {McCarthy}, Donald and {MacQueen}, Phillip and {Meibom}, S{\o}ren and {Miquel}, Thibaut and {Prsa}, Andrej and {Quinn}, Samuel N. and {Quintana}, Elisa V. and {Ragozzine}, Darin and {Sherry}, William and {Shporer}, Avi and {Tenenbaum}, Peter and {Torres}, Guillermo and {Twicken}, Joseph D. and {Van Cleve}, Jeffrey and {Walkowicz}, Lucianne and {Witteborn}, Fred C. and {Still}, Martin},
        title = "{Characteristics of Planetary Candidates Observed by Kepler. II. Analysis of the First Four Months of Data}",
      journal = {\apj},
         year = 2011,
        month = jul,
       volume = {736},
       number = {1},
          eid = {19},
        pages = {19},
          doi = {10.1088/0004-637X/736/1/19},
archivePrefix = {arXiv},
       eprint = {1102.0541},
 primaryClass = {astro-ph.EP},
       adsurl = {https://ui.adsabs.harvard.edu/abs/2011ApJ...736...19B}
}

@ARTICLE{Malamud2024,
       author = {{Malamud}, Uri and {Podolak}, Morris and {Podolak}, Joshua I. and {Bodenheimer}, Peter H.},
        title = "{Uranus and Neptune as methane planets: Producing icy giants from refractory planetesimals}",
      journal = {\icarus},
         year = 2024,
        month = oct,
       volume = {421},
          eid = {116217},
        pages = {116217},
          doi = {10.1016/j.icarus.2024.116217},
       adsurl = {https://ui.adsabs.harvard.edu/abs/2024Icar..42116217M}
}

@ARTICLE{Bergin2023,
       author = {{Bergin}, Edwin A. and {Kempton}, Eliza M. -R. and {Hirschmann}, Marc and {Bastelberger}, Sandra T. and {Teal}, D.~J. and {Blake}, Geoffrey A. and {Ciesla}, Fred J. and {Li}, Jie},
        title = "{Exoplanet Volatile Carbon Content as a Natural Pathway for Haze Formation}",
      journal = {\apjl},
         year = 2023,
        month = may,
       volume = {949},
       number = {1},
          eid = {L17},
        pages = {L17},
          doi = {10.3847/2041-8213/acd377},
archivePrefix = {arXiv},
       eprint = {2305.05056},
 primaryClass = {astro-ph.EP},
       adsurl = {https://ui.adsabs.harvard.edu/abs/2023ApJ...949L..17B}
}

@ARTICLE{Izidoro2021,
       author = {{Izidoro}, Andr{\'e} and {Bitsch}, Bertram and {Raymond}, Sean N. and {Johansen}, Anders and {Morbidelli}, Alessandro and {Lambrechts}, Michiel and {Jacobson}, Seth A.},
        title = "{Formation of planetary systems by pebble accretion and migration. Hot super-Earth systems from breaking compact resonant chains}",
      journal = {\aap},
         year = 2021,
        month = jun,
       volume = {650},
          eid = {A152},
        pages = {A152},
          doi = {10.1051/0004-6361/201935336},
archivePrefix = {arXiv},
       eprint = {1902.08772},
 primaryClass = {astro-ph.EP},
       adsurl = {https://ui.adsabs.harvard.edu/abs/2021A&A...650A.152I}
}

@article{Alexander17,
title = {The nature, origin and modification of insoluble organic matter in chondrites, the major source of Earth’s C and N},
journal = {Geochemistry},
volume = {77},
number = {2},
pages = {227-256},
year = {2017},
issn = {0009-2819},
doi = {https://doi.org/10.1016/j.chemer.2017.01.007},
url = {https://www.sciencedirect.com/science/article/pii/S0009281916301350},
author = {C.M.O’D. Alexander and G.D. Cody and B.T. {De Gregorio} and L.R. Nittler and R.M. Stroud},
}

@ARTICLE{Bitsch2019_waterworld,
       author = {{Bitsch}, Bertram and {Raymond}, Sean N. and {Izidoro}, Andre},
        title = "{Rocky super-Earths or waterworlds: the interplay of planet migration, pebble accretion, and disc evolution}",
      journal = {\aap},
         year = 2019,
        month = apr,
       volume = {624},
          eid = {A109},
        pages = {A109},
          doi = {10.1051/0004-6361/201935007},
archivePrefix = {arXiv},
       eprint = {1903.02488},
 primaryClass = {astro-ph.EP},
       adsurl = {https://ui.adsabs.harvard.edu/abs/2019A&A...624A.109B}
}

@ARTICLE{Madhu12,
       author = {{Madhusudhan}, Nikku and {Lee}, Kanani K.~M. and {Mousis}, Olivier},
        title = "{A Possible Carbon-rich Interior in Super-Earth 55 Cancri e}",
      journal = {\apjl},
         year = 2012,
        month = nov,
       volume = {759},
       number = {2},
          eid = {L40},
        pages = {L40},
          doi = {10.1088/2041-8205/759/2/L40},
archivePrefix = {arXiv},
       eprint = {1210.2720},
 primaryClass = {astro-ph.EP},
       adsurl = {https://ui.adsabs.harvard.edu/abs/2012ApJ...759L..40M}
}

@ARTICLE{Hakim19,
       author = {{Hakim}, Kaustubh and {Spaargaren}, Rob and {Grewal}, Damanveer S. and {Rohrbach}, Arno and {Berndt}, Jasper and {Dominik}, Carsten and {van Westrenen}, Wim},
        title = "{Mineralogy, Structure, and Habitability of Carbon-Enriched Rocky Exoplanets: A Laboratory Approach}",
      journal = {Astrobiology},
         year = 2019,
        month = jul,
       volume = {19},
       number = {7},
        pages = {867-884},
          doi = {10.1089/ast.2018.1930},
archivePrefix = {arXiv},
       eprint = {1807.02064},
 primaryClass = {astro-ph.EP},
       adsurl = {https://ui.adsabs.harvard.edu/abs/2019AsBio..19..867H}
}

@ARTICLE{Rubin19,
       author = {{Rubin}, Martin and {Altwegg}, Kathrin and {Balsiger}, Hans and {Berthelier}, Jean-Jacques and {Combi}, Michael R. and {De Keyser}, Johan and {Drozdovskaya}, Maria and {Fiethe}, Bj{\"o}rn and {Fuselier}, Stephen A. and {Gasc}, S{\'e}bastien and {Gombosi}, Tamas I. and {H{\"a}nni}, Nora and {Hansen}, Kenneth C. and {Mall}, Urs and {R{\`e}me}, Henri and {Schroeder}, Isaac R.~H.~G. and {Schuhmann}, Markus and {S{\'e}mon}, Thierry and {Waite}, Jack H. and {Wampfler}, Susanne F. and {Wurz}, Peter},
        title = "{Elemental and molecular abundances in comet 67P/Churyumov-Gerasimenko}",
      journal = {\mnras},
         year = 2019,
        month = oct,
       volume = {489},
       number = {1},
        pages = {594-607},
          doi = {10.1093/mnras/stz2086},
archivePrefix = {arXiv},
       eprint = {1907.11044},
 primaryClass = {astro-ph.EP},
       adsurl = {https://ui.adsabs.harvard.edu/abs/2019MNRAS.489..594R}
}

@ARTICLE{Li21,
       author = {{Li}, J. and {Bergin}, E.~A. and {Blake}, G.~A. and {Ciesla}, F.~J. and {Hirschmann}, M.~M.},
        title = "{Earth's carbon deficit caused by early loss through irreversible sublimation}",
      journal = {Science Advances},
         year = 2021,
        month = apr,
       volume = {7},
       number = {14},
        pages = {eabd3632},
          doi = {10.1126/sciadv.abd3632},
archivePrefix = {arXiv},
       eprint = {2104.02702},
 primaryClass = {astro-ph.EP},
       adsurl = {https://ui.adsabs.harvard.edu/abs/2021SciA....7.3632L}
}

@ARTICLE{Bitsch2019,
       author = {{Bitsch}, Bertram and {Izidoro}, Andre and {Johansen}, Anders and
         {Raymond}, Sean N. and {Morbidelli}, Alessandro and
         {Lambrechts}, Michiel and {Jacobson}, Seth A.},
        title = "{Formation of planetary systems by pebble accretion and migration: growth of gas giants}",
      journal = {\aap},
         year = 2019,
        month = mar,
       volume = {623},
          eid = {A88},
        pages = {A88},
          doi = {10.1051/0004-6361/201834489},
archivePrefix = {arXiv},
       eprint = {1902.08771},
 primaryClass = {astro-ph.EP},
       adsurl = {https://ui.adsabs.harvard.edu/abs/2019A&A...623A..88B}
}

@article{Johansen2017,
author = { Anders  Johansen and  Michiel  Lambrechts},
title = {Forming Planets via Pebble Accretion},
journal = {Annual Review of Earth and Planetary Sciences},
volume = {45},
number = {1},
pages = {null},
year = {2017},
doi = {10.1146/annurev-earth-063016-020226},

URL = { 
        http://www.annualreviews.org/doi/abs/10.1146/annurev-earth-063016-020226
    
},
eprint = { 
        http://www.annualreviews.org/doi/pdf/10.1146/annurev-earth-063016-020226
    
}
}

@ARTICLE{2016Icar..272...32Q,
       author = {{Quirico}, E. and {Moroz}, L.~V. and {Schmitt}, B. and {Arnold}, G. and {Faure}, M. and {Beck}, P. and {Bonal}, L. and {Ciarniello}, M. and {Capaccioni}, F. and {Filacchione}, G. and {Erard}, S. and {Leyrat}, C. and {Bockel{\'e}e-Morvan}, D. and {Zinzi}, A. and {Palomba}, E. and {Drossart}, P. and {Tosi}, F. and {Capria}, M.~T. and {De Sanctis}, M.~C. and {Raponi}, A. and {Fonti}, S. and {Mancarella}, F. and {Orofino}, V. and {Barucci}, A. and {Blecka}, M.~I. and {Carlson}, R. and {Despan}, D. and {Faure}, A. and {Fornasier}, S. and {Gudipati}, M.~S. and {Longobardo}, A. and {Markus}, K. and {Mennella}, V. and {Merlin}, F. and {Piccioni}, G. and {Rousseau}, B. and {Taylor}, F.},
        title = "{Refractory and semi-volatile organics at the surface of comet 67P/Churyumov-Gerasimenko: Insights from the VIRTIS/Rosetta imaging spectrometer}",
      journal = {\icarus},
         year = 2016,
        month = jul,
       volume = {272},
        pages = {32-47},
          doi = {10.1016/j.icarus.2016.02.028},
       adsurl = {https://ui.adsabs.harvard.edu/abs/2016Icar..272...32Q}
}

@ARTICLE{2018JGRE..123.2295M,
       author = {{Miozzi}, F. and {Morard}, G. and {Antonangeli}, D. and {Clark}, A.~N. and {Mezouar}, M. and {Dorn}, C. and {Rozel}, A. and {Fiquet}, G.},
        title = "{Equation of State of SiC at Extreme Conditions: New Insight Into the Interior of Carbon-Rich Exoplanets}",
      journal = {Journal of Geophysical Research (Planets)},
         year = 2018,
        month = sep,
       volume = {123},
       number = {9},
        pages = {2295-2309},
          doi = {10.1029/2018JE005582},
archivePrefix = {arXiv},
       eprint = {1808.08201},
 primaryClass = {astro-ph.EP},
       adsurl = {https://ui.adsabs.harvard.edu/abs/2018JGRE..123.2295M}
}

@ARTICLE{Mobidelli2000,
       author = {{Morbidelli}, A. and {Chambers}, J. and {Lunine}, J.~I. and {Petit}, J.~M. and {Robert}, F. and {Valsecchi}, G.~B. and {Cyr}, K.~E.},
        title = "{Source regions and time scales for the delivery of water to Earth}",
      journal = {\maps},
         year = 2000,
        month = nov,
       volume = {35},
       number = {6},
        pages = {1309-1320},
          doi = {10.1111/j.1945-5100.2000.tb01518.x},
       adsurl = {https://ui.adsabs.harvard.edu/abs/2000M&PS...35.1309M}
}

@ARTICLE{Kreidberg14,
       author = {{Kreidberg}, Laura and {Bean}, Jacob L. and {D{\'e}sert}, Jean-Michel and
         {Line}, Michael R. and {Fortney}, Jonathan J. and {Madhusudhan}, Nikku and
         {Stevenson}, Kevin B. and {Showman}, Adam P. and {Charbonneau}, David and
         {McCullough}, Peter R. and {Seager}, Sara and {Burrows}, Adam and
         {Henry}, Gregory W. and {Williamson}, Michael and {Kataria}, Tiffany and
         {Homeier}, Derek},
        title = "{A Precise Water Abundance Measurement for the Hot Jupiter WASP-43b}",
      journal = {\apj},
         year = "2014",
        month = "Oct",
       volume = {793},
       number = {2},
          eid = {L27},
        pages = {L27},
          doi = {10.1088/2041-8205/793/2/L27},
archivePrefix = {arXiv},
       eprint = {1410.2255},
 primaryClass = {astro-ph.EP},
       adsurl = {https://ui.adsabs.harvard.edu/abs/2014ApJ...793L..27K}
}

@ARTICLE{Oberg11_C_O,
   author = {{{\"O}berg}, K.~I. and {Murray-Clay}, R. and {Bergin}, E.~A.
	},
    title = "{The Effects of Snowlines on C/O in Planetary Atmospheres}",
  journal = {ApJL},
archivePrefix = "arXiv",
   eprint = {1110.5567},
 primaryClass = "astro-ph.GA",
     year = 2011,
    month = dec,
   volume = 743,
      eid = {L16},
    pages = {L16},
      doi = {10.1088/2041-8205/743/1/L16},
   adsurl = {http://adsabs.harvard.edu/abs/2011ApJ...743L..16O}
}

@ARTICLE{Gail17,
   author = {{Gail}, H.-P. and {Trieloff}, M.},
    title = "{Spatial distribution of carbon dust in the early solar nebula and the carbon content of planetesimals}",
  journal = {\aap},
archivePrefix = "arXiv",
   eprint = {1707.07611},
 primaryClass = "astro-ph.EP",
     year = 2017,
    month = sep,
   volume = 606,
      eid = {A16},
    pages = {A16},
      doi = {10.1051/0004-6361/201730480},
   adsurl = {http://adsabs.harvard.edu/abs/2017A%26A...606A..16G}
}

@INPROCEEDINGS{Bond11,
   author = {{Bond}, J.~C. and {O'Brien}, D.~P. and {Raymond}, S.~N.},
    title = "{The Diversity of Extrasolar Terrestrial Planets: adding
migration into the mix.}",
booktitle = {EPSC-DPS Joint Meeting 2011},
     year = 2011,
    month = oct,
    pages = {160},
   adsurl = {http://adsabs.harvard.edu/abs/2011epsc.conf..160B}
}

@article{Bergin15,
	Adsurl = {http://adsabs.harvard.edu/abs/2015PNAS..112.8965B},
	Archiveprefix = {arXiv},
	Author = {{Bergin}, E.~A. and {Blake}, G.~A. and {Ciesla}, F. and {Hirschmann}, M.~M. and {Li}, J.},
	Doi = {10.1073/pnas.1500954112},
	Eprint = {1507.04756},
	Journal = {Proceedings of the National Academy of Science},
	Month = jul,
	Pages = {8965-8970},
	Primaryclass = {astro-ph.EP},
	Title = {{Tracing the ingredients for a habitable earth from interstellar space through planet formation}},
	Volume = 112,
	Year = 2015}

@ARTICLE{2014ApJS..215...21B,
       author = {{Becker}, Andreas and {Lorenzen}, Winfried and {Fortney}, Jonathan J. and {Nettelmann}, Nadine and {Sch{\"o}ttler}, Manuel and {Redmer}, Ronald},
        title = "{Ab Initio Equations of State for Hydrogen (H-REOS.3) and Helium (He-REOS.3) and their Implications for the Interior of Brown Dwarfs}",
      journal = {\apjs},
         year = 2014,
        month = dec,
       volume = {215},
       number = {2},
          eid = {21},
        pages = {21},
          doi = {10.1088/0067-0049/215/2/21},
archivePrefix = {arXiv},
       eprint = {1411.4010},
 primaryClass = {astro-ph.EP},
       adsurl = {https://ui.adsabs.harvard.edu/abs/2014ApJS..215...21B}
}

@ARTICLE{2009PhRvB..79e4107F,
       author = {{French}, Martin and {Mattsson}, Thomas R. and {Nettelmann}, Nadine and {Redmer}, Ronald},
        title = "{Equation of state and phase diagram of water at ultrahigh pressures as in planetary interiors}",
      journal = {\prb},
         year = 2009,
        month = feb,
       volume = {79},
       number = {5},
          eid = {054107},
        pages = {054107},
          doi = {10.1103/PhysRevB.79.054107},
       adsurl = {https://ui.adsabs.harvard.edu/abs/2009PhRvB..79e4107F}
}

@article{Armstrong15,
	Author = {{Armstrong}, L.~S. and {Hirschmann}, M.~M. and {Stanley}, B.~D. and {Falksen}, E.~G. and {Jacobsen}, S.~D.},
	Journal = {\gca},
	Pages = {submitted},
	Title = {{Speciation and solubility of reduced C-O-H-N volatiles in basaltic melt: Implications for volcanism, atmospheric evolution, and deep volatile cycles in the terrestrial planets}},
	Year = 2015}

@ARTICLE{Luque22,
       author = {{Luque}, Rafael and {Pall{\'e}}, Enric},
        title = "{Density, not radius, separates rocky and water-rich small planets orbiting M dwarf stars}",
      journal = {Science},
         year = 2022,
        month = sep,
       volume = {377},
       number = {6611},
        pages = {1211-1214},
          doi = {10.1126/science.abl7164},
archivePrefix = {arXiv},
       eprint = {2209.03871},
 primaryClass = {astro-ph.EP},
       adsurl = {https://ui.adsabs.harvard.edu/abs/2022Sci...377.1211L}}

@article{Sephton03,
	Adsurl = {http://adsabs.harvard.edu/abs/2003GeCoA..67.2093S},
	Author = {{Sephton}, M.~A. and {Verchovsky}, A.~B. and {Bland}, P.~A. and {Gilmour}, I. and {Grady}, M.~M. and {Wright}, I.~P.},
	Doi = {10.1016/S0016-7037(02)01320-0},
	Journal = {\gca},
	Month = jun,
	Pages = {2093-2108},
	Title = {{Investigating the variations in carbon and nitrogen isotopes in carbonaceous chondrites}},
	Volume = 67,
	Year = 2003}

@article{Ardia13,
	Adsurl = {http://adsabs.harvard.edu/abs/2013GeCoA.114...52A},
	Author = {{Ardia}, P. and {Hirschmann}, M.~M. and {Withers}, A.~C. and {Stanley}, B.~D.},
	Doi = {10.1016/j.gca.2013.03.028},
	Journal = {\gca},
	Month = aug,
	Pages = {52-71},
	Title = {{Solubility of CH$_{4}$ in a synthetic basaltic melt, with applications to atmosphere-magma ocean-core partitioning of volatiles.}},
	Volume = 114,
	Year = 2013}
\bibliographystyle{aasjournalv7}

%The corresponding author is responsible for submitting a \href{http://www.nature.com/srep/policies/index.html#competing}{competing interests statement} on behalf of all authors of the paper. This statement must be included in the submitted article file.

\newpage
\appendix
This study investigates the potential role of carbon-rich compounds in the interiors of exoplanets. There are considerable uncertainties in the internal structure and the thermal profiles of the modeled planets as well as in the distribution and forms of the soot component in exoplanets. We derive bounds on the mass-radius relations by treating the soot component as a fictive phase contributing to the properties of the planets, and then estimating the density of the soot component on the basis of a broad correlation between density and mean atomic number. We found that the uncertainties in the forms of carbon in exoplanets and the equations-of state (EoS) of relevant phases are comparable to the observational uncertainties in mass and radius; therefore, our simplifying approach allows meaningful estimates to examine whether soot planets can account for observed mass-radius relations. \\
 
\section{Model Planet Compositions} 
Guided by the solar, meteoritic, and cometary compositions, together with first-order constraints from planet formation models \citep{Li21}, we constructed model planets from four chemical components: rock, soot, water, and hydrogen (Table~\ref{Tab:composition}). The rock component contains four elements, oxygen (O), magnesium (Mg), silicon (Si), and iron (Fe), with the solar Mg/Si and Fe/Si ratios \citep{Lodders10}. It consists of two phases, with all Mg and Si combined with O to form MgSiO$_3$, and all Fe present as metal. The soot component is assumed to have C:H:O atomic ratio of 100:77±2:14±3, the average value of insoluble organic matter (IOM) in primitive chondrites and comet Halley \citep{Alexander17}. It consists of multiple phases, including meteoritic insoluble organic matter (IOM) and carbonaceous materials found in comets. The water component and hydrogen component are taken to be  pure H$_2$O and pure H$_2$, respectively. The specific carriers of the carbon are a variety of materials, including species identified above, which we term “refractory organics.” 

The assumed C/Si ratio is consistent with the observation that on average dust is half rock and half soot \citep{Rubin19} and is also consistent with C/Si ratios detected in some polluted white dwarfs  \citep{koester2014atmospheric, 2024A&A...691A.352W}, in comets 67P \citep{Bardyn2017}, Halley \citep{Fomenkova1999}, and other comets \citep{Woodward2021}. The C/H ratios vary from 0.1-0.2 in Murchison meteorites, 0.3-0.5 in Bennu, and 0.4 in Tagish Lake \citep{2025NatAs...9..199G}, to 1.7 in Ryugu\citep{2025GeocJ..59...45Y}. We adopted the C/H ratio of 1.3 for IOM \citep{Alexander17}. This ratio affects the relative amounts of soot and water in the models of soot planets and soot-water worlds, but does not change our conclusions.
 
We consider three types of model planets with four distinct chemical compositions: a rocky planet, a soot planet, a dry version of soot-water world, and a wet version of soot-water world(Table~\ref{Tab:composition}). The model rocky planet only contains the rock component. The model soot planet contains the rock and soot components and has a C/Si ratio at half of the solar value \citep{Lodders10}, following the assumption that half of the solar carbon is preserved as refractory organics and the other half is lost as carbon monoxide (CO) \citep{Li21}. This assumption is consistent with the observation that on average dust is half rock and half soot\citep {Rubin19} and is also consistent with C/Si ratios detected in some polluted white dwarfs \citep{koester2014atmospheric, 2024A&A...691A.352W}, in comets 67P \citep{Bardyn2017}, Halley \citep{Fomenkova1999}, and numerous others \citep{Woodward2021}. Hirschmann et al. \citep{Hirschmann2021PNAS..11826779H} showed that devolatilization during small-body differentiation is a key process in shaping the volatile inventory of terrestrial planets derived from planetesimals and planetary embryos. We thus consider a range of dust-to-ice ratio to account for variable degrees of volatile loss by parent-body processing during the formative stages of planets. For the soot-water worlds, we consider 1 as a lower bound on the dust-to-ice ratio and 3 as an upper bound, as observed in Comet 67P \citep{Rubin19, Malamud2024}\citep{2015Icar..246...21M,  2019Icar..326...10B}. Accordingly, the model soot planet contains about one quarter of soot and three quarters of rock by mass, and the model dry and wet soot-water worlds consist of about $\frac{1}{2}$rock-$\frac{1}{5}$soot-$\frac{1}{4}$water and $\frac{1}{3}$rock-$\frac{1}{7}$soot-$\frac{1}{2}$water, respectively (Table~\ref{Tab:composition}).
% Some Fe in the rock component is likely combined with O to form FeO and Fe$_2$O$_3$, as in Earth's mantle. Uncertainties in CHO ratio of soot leads to ±0.5\% in the mass fractions. 
\setcounter{table}{0}
\renewcommand{\thetable}{A\arabic{table}}
\setcounter{figure}{0}
\renewcommand{\thefigure}{A\arabic{figure}}

%(N$_2$)$_{0.99}$(CH$_4$)$_{0.01}$, $\rho_0$ = 0.95$\pm$0.11, Horst13; Tholin produced by UV irradiation of N$_2$ with 1\% CH$_4$.}
The compositions of real planets likely deviate from these models. For instance, inclusion of hydrous minerals may alter a planet's H/C ratio, as H$_2$O can be stored in hydrous minerals such as clays and serpentine in chondritic meteorites and asteroids (e.g.\citep{Bitsch2019,2019NatAs...3..332H}). Formation of hydrous phases requires reaction between anhydrous precursors and water ice and therefore hydrous phases are abundant only beyond water snow line. Once formed, hydrous phases may be delivered to the inner disks through radial migration and they may survive the high temperatures due to sluggish dehydration \citep{Ciesla2005E&PSL.231....1C}. Incorporating hydrous minerals would increase the H/C ratios of rocky or soot planets.  On the other hand, hydrous phases are thought to form via parent body alteration, during which water loss is expected, while soot is expected to survive the process, which occurs below the decomposition temperatures of the soot. As a result, the presence of hydrous minerals in soot-water worlds may be associated with lower H/C ratios.  Finally, the required parent body processing also implies that a planet accreting from pebbles likely contains little hydrous materials.

\begin{table}[h]
\centering
\caption{Compositions of rocky planet, soot planet, and soot-water worlds}
\label{Tab:composition}
\begin{tabular}{lcccc}
\hline
 &\textbf{Rocky~Planet}
    &\textbf{Soot~Planet}
    &\parbox{4cm}{\centering\textbf{Soot-Water World} \\ \textbf{(dry)}}
    &\parbox{4cm}{\centering\textbf{Soot-Water World} \\ \textbf{(wet)}} \\
\hline
\multicolumn{5}{l}{\textbf{Element (atomic ratio)}$^{1}$} \\
H  & --    & 2.76 & 7.50 & 11.84 \\
C  & --    & 3.54 & 3.54 & 3.54 \\
Mg & 1.02  & 1.02 & 1.02 & 1.02 \\
Si & 1     & 1    & 1    & 1    \\
Fe & 0.838 & 0.838 & 0.838 & 0.838 \\
\hline
\multicolumn{5}{l}{\textbf{Component (mass\%)}$^{2}$} \\
Rock  & 100     & 74(5) & 55(1) & 36(1) \\
Soot  & --      & 26(5) & 20(1) & 14(1) \\
H$_2$O & --     & --    & 25(1) & 50(1) \\
\vspace{-5pt}
\\
Iron   & 32(1)  & 24(1) & 18(1) & 13(1) \\
Silicate & 68(1) & 50(1) & 37(1) & 24(1) \\
\hline
\multicolumn{5}{l}{\footnotesize $^{1}$ With Si fixed at 1 \cite{Lodders03} ``--'' denotes negligible.} \\
\multicolumn{5}{l}{\footnotesize $^{2}$ Volume fractions for Earth-based planet ($R$ = $R_e$ only).} \\
\end{tabular}
\end{table}

Beyond the water snow line, solids are expected to contain about 50\% water. In meteorites where hydrous phases are found, the water contents are often well below 10\% by mass. The relatively water-poor nature of these meteorites suggest limited incorporation of hydrous phases inside the water snow line, extensive water loss from the acquired hydrous phases, and/or pebble accretion as the dominant formation mechanism.

Given these considerations, we expect that the inclusion of hydrous minerals does not substantially raise the H/C ratios of rocky or soot planets, whereas the presence of hydrous minerals in soot-water worlds likely lowers their H/C ratios.

Note that the previous water worlds \citep{Luque22} consist of rock and water, whereas our model soot-water worlds contain soot as a major component, as it is unlikely that a planet can accrete appreciable water without also accreting soot \citep{Li21}. Carbon-rich compounds are known to significantly contribute to the mass budget of a variety of objects in the Solar System including comets, carbonaceous chondrites, icy moons, TNOs, and ice giants, so they should contribute to the composition of planets around other stars, so long as they coalesced outside the “soot line”. Combining chemical data from materials in the solar system (e.g., comets and chondritic meteorites) and thermochemical modeling, our study demonstrates that water worlds are also expected to contain soot, and that planets rich in “soot” (refractory organic carbon) can account for the observed exoplanets with low average densities that were previously attributed to a binary combination of rock and water.

Because soot consists largely of hydrocarbons, the interiors of planets rich in soot will also be comparatively reduced and produce secondary atmospheres rich in methane \citep{Bergin2023}.  {Reactions such as 6 CH (in soot) + Fe$_2$O$_3$ = C + 2FeO + 3H$_2$O will effectively neutralize any oxidized iron so long as the soot/oxidized iron mass ratio is greater than about 0.3. In soot planets the total soot mass is more than 20\% and the majority of iron is accreted as metal, and therefore volcanogenic C is outgassed as a combination of CH$_4$ and CO.

\section{C/O Ratio}
There is a growing literature on the overall carbon to oxygen content of exoplanetary atmospheres along side their potential link to the disk composition, all in the context of an assumed/known stellar C/O ratio \citep{Oberg11_C_O, 2012ApJ...758...36M, 2016ApJ...817...17G,2013ApJ...763...25M, 2019ARA&A..57..617M, 2024ApJ...969L..21B}.
The overall context is described by \citep{Oberg11_C_O} where the stellar C/O is assumed to be solar and the disk composition reflects that the major carriers of elemental C and O based on ISM constraints.  These are silicate dust, H$_2$O, CO, and CO$_2$ for oxygen and carbonaceous dust, CO, and CO$_2$ for carbon.  This particular model mostly focused on the volatile ice components that condense $\ge$1~au from the star (i.e. CO, CO$_2$, and H$_2$O).  This has relevance for the composition of {\em gas giants} in the context of the core-accretion model of giant planet formation as discussed in the references above.

The model discussed in this paper is an extension of this concept, but with a focus on the innermost region of protoplanetary disks from the water ice line and inwards.  This includes the sublimation fronts of soot (carbonaceous dust) \citep{Kress2010, Li21} and silicates.  It is this region that matters most for rocky planets (or the cores of mini-Neptune's) as discussed by \citet{Li21} and \citet{Bergin2023}. Additional motivational work is found in Bond et al. \citep{Bond11} and Moriarty et al.\citep{2014ApJ...787...81M} who use thermochemical equilibrium simulations to argue that carbon-rich bodies could be present in the extra-solar inventory \citep[see also][]{2025AJ....169..180S}. However, the formation of macromolecular hydrocarbons is not a product of equilibrium condensation \citep{Lodders03}.  Thus, the concept of the irreversible destruction of soot via sublimation \citep{Li21} places a new constraint on the location of carbon-rich material with attendant implications for planet formation. 

In this regard, a central question is how common is the C/O ratio of the solar system in comparison to exoplanetary systems. \citet{Madhu12} explored the possibility of a carbon-rich interior for the super-Earth 55 Cancri under the assumption of a stellar C/O $>$ 1.  This composition directly implies the presence of excess carbon in the disk.  If this excess carbon could be placed in refractory solids, then this would potentially be reflected in the composition of the rocky planet \citep{Johansen2017}.   In Fig.~\ref{fig:star_c_o} we present a compilation of stellar C/O ratios for stars that host known planetary systems with massive planets (M $\ge$30~M$_\oplus$; shown in Red) and low mass planets (shown in Blue).  As can be seen here, for both cases, the vast majority of systems have near solar C/O ratios.  This is consistent with other analysis \citep{Nissen2013, 2024A&A...688A.193D} and the argument that systems with extreme C/O ratios ($>$ 1) are rare in the nearby stellar population \citep{2012ApJ...747L..27F}.

\begin{figure}[h]
\centering
\includegraphics[width=0.6\linewidth]{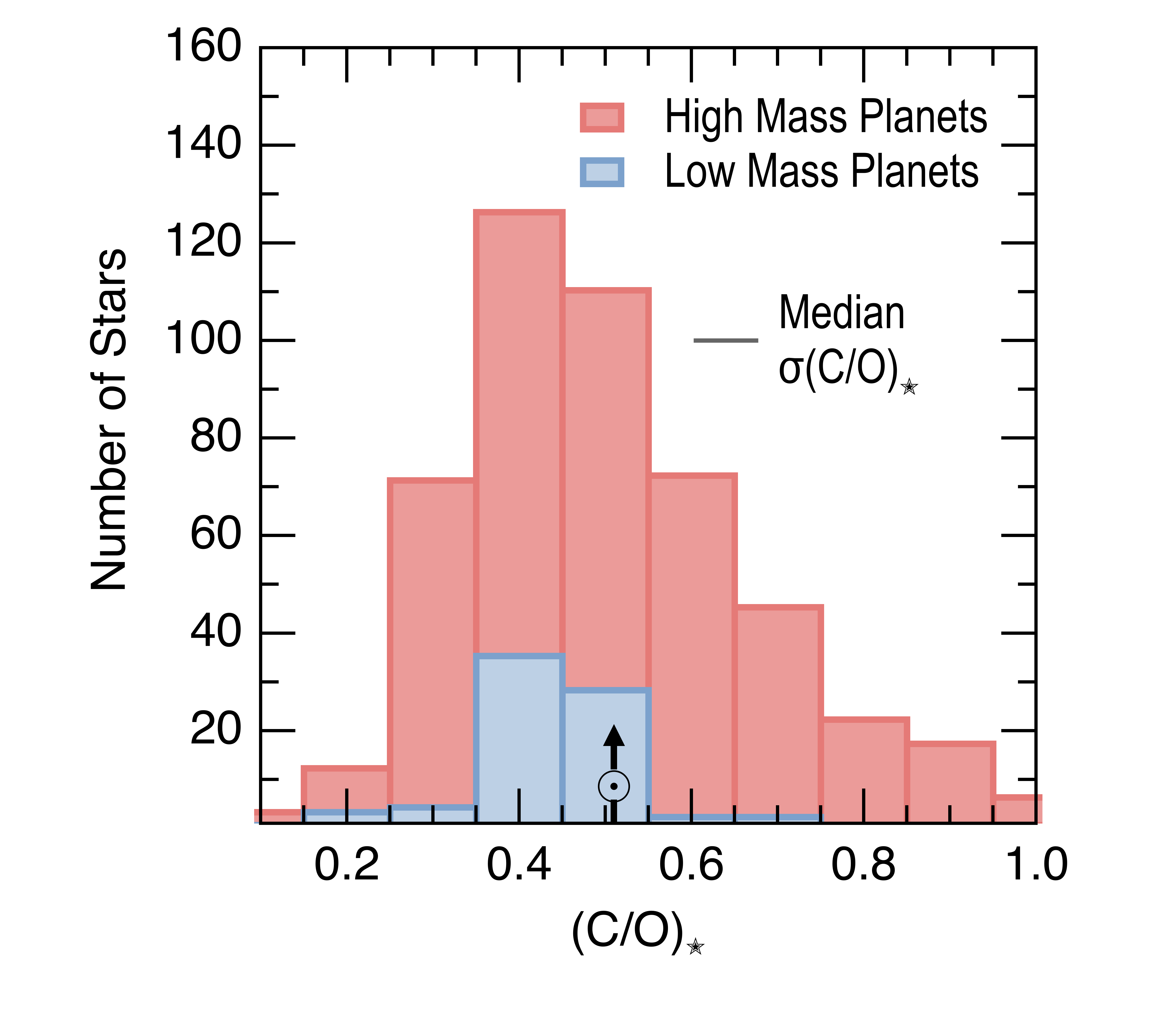}
\caption{Histogram of stellar C/O ratios, (C/O)$_\star$, for planet host stars as compiled by the Hypatia Catalog \citep{Hypatia}. The histogram shown in red (High Mass Planets) isolates systems where the most massive planet is $\ge$30~M$_\oplus$ and the one is blue (Low Mass Planets) isolates those with all planet mass below that value. The median error for each sample is shown with its magnitude comparable to the length of the line.  The solar value is also show as reference using oxygen abundance estimates by Bergemann et al. \protect\citep{2021MNRAS.508.2236B} and oxygen by Asplund et al. \protect\citep{2021A&A...653A.141A}.}
\label{fig:star_c_o}
\end{figure}

The model presented in this paper assumes solar composition and that material beyond the soot and water ice lines is organic rich, based on the composition of comets in our solar system\citep{Fomenkova99, Rubin19, 2021PSJ.....2...25W}.  In astronomical observation, it is inferred that organic-rich carbonaceous grains is present in almost every model of disk dust emission from 0.1~$\mu$m to cm wavelengths \citep{1998ApJ...500..411D, 2001ApJ...553..321D, 2005ApJ...631.1134A, 2010A&A...512A..15R, 2018ApJ...869L..46D}, as carbon grains are major sources of opacity \citep{1994ApJ...421..615P, 1996MNRAS.282.1321Z, 2018ApJ...869L..45B}.  These models assume that carbon-rich organic material contains $\sim$ 50\% elemental carbon on the basis of cometary and meteorite composition \citep{1994ApJ...421..615P} and models of interstellar grain extinction \citep{Jones13, Chiar13}.

In sum, the solar system and analyses of astronomical observation infer a significant amount of carbon in refractory solids.  Beyond the soot line this material is a major component of their composition and must be taken into account.

\section{Soot Density}
To estimate the density of the soot component at ambient pressure and temperature conditions (100 kPa = 1 bar, 300 K), we compiled the densities of an array of planet-forming materials, including iron metal, silicates, water ice, and various carbon-bearing phases including diamond, graphite, carbonates, and iron carbides (Table~\ref{Tab:rho_z}). The compilation also includes a variety of ices observed at cryogenic conditions as well metastable phases which form at the high pressures prevailing in planetary interiors. Due to thermal expansion upon heating from 80 K to 195 K, the densities of H$_2$O\textsuperscript{1h} ice and CO$_2$\textsuperscript{I} ice decrease by 0.7\% and 8\%, respectively \citep{Mangan2017}. At 300 K, the densities of fictitious H$_2$O\textsuperscript{1h} ice and CO$_2$\textsuperscript{I} ice are 0.896 g/cc and 1.398 g/cc, respectively, after correcting for thermal expansion.
%\clearpage
%\newpage
%\thispagestyle{empty}

\begin{table}[h]
\caption{Densities and Mean Atomic Numbers of 48 Carbon-Bearing Phases}
\label{Tab:rho_z}
\hspace*{-4em}
\begin{tabular}{l c c c c}
\hline
Phase& $\overline{Z}$ & $\rho_0$\textsuperscript{a}, g/cm\textsuperscript{3} & $T$\textsuperscript{b}, K & Ref.\\
\hline
CH$_4$	&2.00	&0.508(5)	&30 &[1]\\
C$_2$H$_6$	&2.25	&0.508(5)	&60 &[1]\\
C$_3$H$_8$	&2.36	&0.778(5)	&65 &[1]\\
NH$_3$	&2.50	&0.797(5)	&95 &[1]\\
CH$_3$NH$_2$	&2.57	&0.834(5)	&100 &[1]\\

C$_2$H$_5$NH$_2$	&2.60	&0.924(5)	&100 &[1]\\
c-C$_3$H$_6$	&2.67	&0.916(5)	&65 &[1]\\
C$_2$H$_4$	&2.67	&0.796(5)	&60 &[1]\\
C$_2$H$_5$OH	&2.89	&0.989(5)	& 120 &[1]\\
CH$_3$OH	&3.00	&1.023(5)	&120 &[1]\\

(CH$_4$)$_4$(H$_2$O)$_{23}$	&3.03	&0.912(5)	& 273 &[2]\\
(CH$_3$)$_2$CO	&3.20	&0.999(5)	&125 &[1]\\
C$_2$H$_5$CN	&3.33	&0.992(5)	&110 &[1]\\
H$_2$O\textsuperscript{Ih} &3.33 &0.917(3) &273 &[3]\\
H$_2$O\textsuperscript{VII}	&3.33 &	1.24(12) & & [4]\\

CH$_3$O	&3.40	&0.970(5)	&75 &[]\\
c-OC$_2$H$_4$	&3.43	&1.142(5)	&100 &[1]\\
HC(O)CH$_3$	&3.43	&1.111(5)	&100 &[1]\\
CH$_4$CN	&3.43	&1.108(5)	& 125 &[1]\\
C$_6$H$_6$	&3.50	&1.085(5)	& 100 &[1]\\

CH$_3$COOCH$_3$	&3.64	&1.197(5)	& 115 &[1]\\
CH$_3$CN	&3.67	&1.073(5)	& 130 &[1]\\
C$_5$H$_5$N	&3.82	&1.149(5)	& 120 &[1]
\vspace{6pt}\\

\textbf{Soot}\textsuperscript{c}	&\textbf{4.17} &\textbf{1.32(5)} & &
\vspace{6pt} \\	

\hline
\end{tabular}
%\hfill
\hspace*{1em}
\begin{tabular}{l c c c c}
\hline
Phase &$\overline{Z}$ & $\rho_0$, g/cm\textsuperscript{3} & $T$, K & Ref.\\
\hline
CH$_3$COOH	&4.00	&1.268(5)	&120 &[1]\\
HCN	&4.67	&1.037(5)	&120  &[1]\\
H$_2$S	&6.00	&1.224(5)	&80 &[1]\\
\textbf{C\textsuperscript{graphite}}	&6.00	&2.281(5) & &[5]\\
C\textsuperscript{diamond}	&6.00	&3.516(5)	& &[5]\\

CO$^\alpha$	&7.00	&1.029(5)  &30 &[6]\\	
N$_2$	&7.00	&1.013(5)	&19 &[1]\\
N$_2$O	&7.33	&1.591(5)	&70 &[1]\\
CO$_2$\textsuperscript{I}	 &7.33	&1.565(3)	&195 &[7]\\
\textbf{Mg$_3$Si$_2$O$_5$(OH)$_4$}	&7.80	&2.625(5)	& &[5]\\

\textbf{MgCO$_3$}	&8.40	&3.010(5) & &[5]\\
\textbf{MgSiO$_3$\textsuperscript{enstatite}} &8.80	&3.204(5) &  &[5]\\
MgSiO$_3$\textsuperscript{bridgmanite}	&8.80	&4.176(5)	&   &[8]\\
OCS	&10.00	&1.518(5)	& 120 &[1]\\
\textbf{Mg$_2$SiO$_4$\textsuperscript{olivine}}	&10.00	&3.227(5)	& &[5]\\

\textbf{SiC\textsuperscript{ZnS}} &10.00 &3.210(5)  & &[9]\\
\textbf{CaCO$_3$}	&10.00	&2.711(5) & &[5]\\
SO$_2$	 &10.67	&1.893(5)	& 85   &[1]\\
\textbf{FeCO$_3$}	&11.20	&3.937(5)  & &[5]\\
\textbf{Fe$_2$SiO$_4$}	&14.00	&4.402(5)	& &[5]\\

Fe$_7$C$_3$	&20.00	&7.680(5) & &[10]\\
\textbf{Fe$_3$C}	&21.00	&7.674(5)	& &[11]\\
\textbf{Fe\textsuperscript{bcc}}	&26.00	&7.873(5) & &[5]\\
Fe\textsuperscript{fcc}	&26.00	&8.170(5) & &[12]\\
Fe\textsuperscript{hcp}	&26.00	&8.430(5) & &[13]\\ 
\hline
\end{tabular}

{\raggedright
\textsuperscript{a}Density at 100 kPa. Values in parentheses are uncertainties on the last digit (assumed to be 5 if not reported).~~  
\textsuperscript{b}Temperature is 298 K if not listed.~~\textsuperscript{c}C:H:O = 100:78:17 in atomic ratio.~~The 11 phases in bold are thermodynamically stable at ambient conditions, while other phases are stable at high pressures or cryogenic temperatures.~~References: [1] Yarnall et al. (2022); [2] Gabitto \& Tsouris (2010); [3] Feistel \& Wagner (2006); [4] Prakapenka et al. (2021); [5] Smyth \& McCormick (1995); [6] Gerakines et al. (2023); [7] Mangan et al. (2017); [8] Fei et al. (2021); [9] Harris (1995); [10] Chen et al. (2012); [11] Li et al. (2002); [12] Komabayashi \& Fei (2010); [13] Smith et al. (2018).}
\end{table}

The densities of the selected phases at 100 kPa and 300 K show a linear correlation with their mean atomic numbers ($\overline{Z}$) (Fig.~\ref{fig:rho_z}). As expected, high-pressure polymorphs such as water ice VII, diamond, bridgmanite, and Fe$_7$C$_3$ iron carbide plot above the linear trend. Most cryogenic ices and simple organic materials follow the linear correlation, with a few outliers falling below the trend (Fig.~\ref{fig:rho_z} inset). A least-squares fit of eleven stable phases at ambient conditions (Table~\ref{Tab:rho_z}) yields $\rho_0$ = 0.327 $\overline{Z}$, with R$^2$ = 0.99, or $\rho_0$ = 0.316 $\overline{Z}$ + 0.161, with R$^2$ = 0.95.  The $\overline{Z}$ of soot with {C:H:O = 100:78:18} in atomic ratio is 4.17 and has a estimated density of 1.32(6) g/cc at 100 kPa and 300 K based on the adopted regression of $\rho_0$ = 0.317(15) $\overline{Z}$. This value is comparable to the proposed density of carbonaceous matter (1.4 g/cc) that coexists with hydrous silicates at 100 kPa and up to 900 K \citep{neri2020}.

To test the validity of our fit, we applied linear fits to the 43 phases that are thermodynamically stable at 100 kPa and 300 K, and to all 48 phases (Fig.\ref{fig:rho_z}). Furthermore, we performed Bayesian analysis, by modeling the data as:
\[
\rho_{\text{model}}(Z) = a Z^b
\]
where $a > 0$ and $b$ are parameters to be inferred.

The prior distributions are:
\[
a \sim \mathcal{U}(0, 20)
\]
\[
b \sim \mathcal{U}(-5, 5)
\]
where $\mathcal{U}(x, y)$ denotes a uniform distribution between $x$ and $y$.

The likelihood for each observation $(Z_i, \rho_i)$ with known uncertainty $\sigma_{\rho, i}$ is:
\[
\rho_{\text{obs}, i} \sim \mathcal{N}(\rho_{\text{model}}(Z_i), \sigma_{\rho, i})
\]
or explicitly:
\[
\mathcal{L}(\{a, b\} \mid \{\rho_i, Z_i\}) = \prod_{i=1}^{N} \frac{1}{\sqrt{2\pi} \sigma_{\rho, i}} \exp\left( -\frac{[\rho_i - a Z_i^b]^2}{2\sigma_{\rho, i}^2} \right)
\]

The posterior is proportional to the product of the likelihood and the priors:
\[
P(a, b \mid \text{data}) \propto \mathcal{L}(\{a, b\} \mid \{\rho_i, Z_i\}) \times P(a) \times P(b)
\]
\[a = 0.232(1)\
 b = 1.097(2)
\]
The results of Bayesian analysis (Fig.\ref{fig:bayesian}) support the estimated soot density of 1.32(6) g/cc.

\begin{figure}[ht]
       \centering
    \begin{minipage}{0.45\textwidth}
        \centering
        \includegraphics[width=\linewidth]{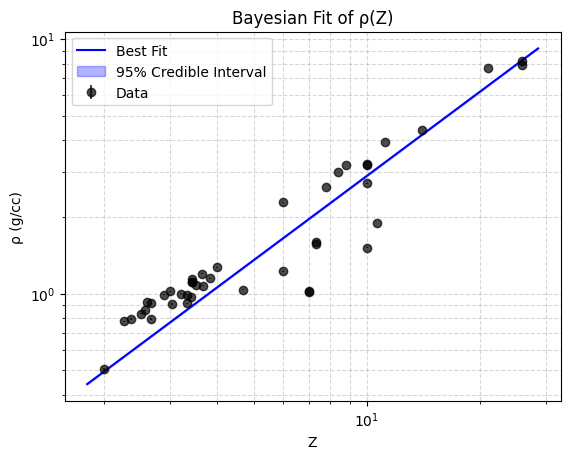}
    \end{minipage}
    \hspace{0.05\textwidth}  % adjustable space between
    \begin{minipage}{0.45\textwidth}
        \centering
        \includegraphics[width=\linewidth]{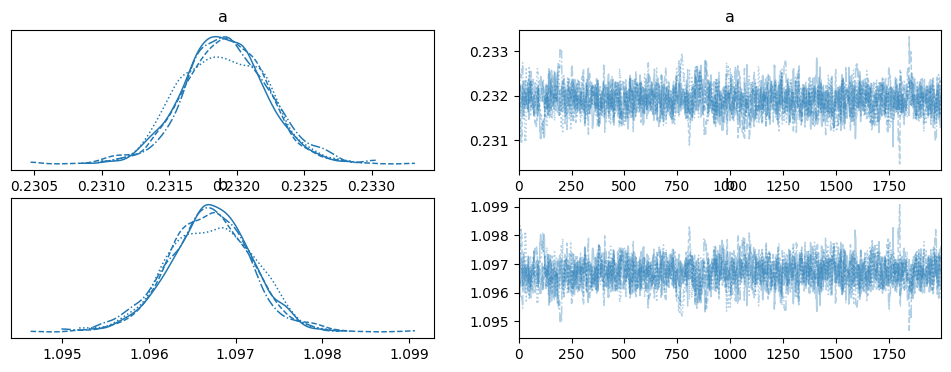}
    \end{minipage}
    \caption{Correlation between density and mean atomic number at 100 kPa (1 bar).  \label{fig:bayesian}}
   \end{figure}

%measurement uncertainties are mostly 0.005, sometimes 0.003 or 0.012
%\newpage
\section{M-R Relations}
We calculated the mass-radius (M-R) relations of model planets through a finite-element iterative algorithm, assuming the Rose-Vinet form of equation-of-state for all phases at 300 K, except hydrogen \citep{Vinet1989}, 
\[P = 3K_0X^{-2}(1-X)e^{(1.5K'-1.5)(1-X)}\]
\[X = \frac{V}{V_0}^\frac{1}{3}\]
where $V_0$ is the volume, $K_0$ is the bulk modulus at 100 kPa, and $K'$ is the first pressure-derivative of the bulk modulus, all at 100 kPar, ad $V$ is the volume at pressure $P$.

A multi-layer planet consists of four distinct layers – an iron core, a silicate mantle, a soot layer, and an outer water layer.  A single-layer planet is modeled as a uniform mixture of iron, silicate, soot, and ice, using the Voigt-Reuss-Hill average:
\begin{align*}
\rho_v &= \sum_i m_i \rho_i \\
\rho_r &= \frac{1}{\sum_i \frac{m_i}{\rho_i}} \\
K_v &= \sum_i m_i K_i \\
K_r &= \frac{1}{\sum_i \frac{m_i}{K_i}} \\
Kp_v &= \sum_i m_i Kp_i \\
Kp_r &= \frac{1}{\sum_i \frac{m_i}{Kp_i}} \\
\\
\rho &= \tfrac{1}{2} (\rho_v + \rho_r) \\
K &= \tfrac{1}{2} (K_v + K_r) \\
Kp &= \tfrac{1}{2} (Kp_v + Kp_r)
\end{align*}
where $i$ is the index over components, $m_i$ is the mass fraction of component $i$ (with $\sum_i m_i = 1$), $\rho_i$ is the density of component $i$, $K_i$ is the bulk modulus of component $i$, $Kp_i$ is the pressure derivative of the bulk modulus of component $i$, $\sum_i$ denotes summation over all components, $\rho_v$ is the Voigt average density (arithmetic mean), $\rho_r$ is the Reuss average density (harmonic mean), $K_v$ is the Voigt average bulk modulus, $K_r$ is the Reuss average bulk modulus, $Kp_v$ is the Voigt average of $Kp$, $Kp_r$ is the Reuss average of $Kp$, $\rho$ is the Hill averaged density, $K$ is the Hill averaged bulk modulus, $Kp$ is the Hill averaged pressure derivative, and $\tfrac{1}{2}$ indicates taking the arithmetic mean of the Voigt and Reuss bounds. We note this is an approximation as a homogeneous phase formed by chemically miscible components is not equivalent to a physical mixture of distinct phases. 

The thickness of a hydrogen envelope surrounding a rocky planet is calculated using a fitted polynomial expression for hydrogen across gaseous and solid states \citep{Tkacz2002}, instead of a Vinet EoS for crystalline hydrogen \citep[e.g.,][]{Loubeyre1996} that is only applicable at pressures above $\sim$10 GPa.

\begin{table}[h]
\caption{Equation-of-State Parameters}
\label{Tab:eos}
\hspace*{2em}
  \begin{tabular}{l l l l l l}
  
\hline  
&Fe\textsuperscript{hcp} [1] &MgSiO$_{3}$\textsuperscript{bdg} [2] &C\textsuperscript{dia}[3] &H$_{2}$O\textsuperscript{ice} [4] & \\ 
\hline 
$\rho_0$\textsuperscript{a}, g/cm\textsuperscript{3} &8.43 &4.176   &3.511 &1.36(2) &\\
[0.5ex]
$\rho_0$\textsuperscript{b} &7.87 &3.22(1) &2.262(3)  &1.36(2)&\\
$K_0$, GPa &177.7(6)  &265.5 &438 &7.9(5)&\\
$K'$ &5.64(1)  &4.16 &3.68 &7.4(2)&\\
$\alpha_0, 10^{-5}$/K &45 &20 &20 &57 & \\
$\gamma_0$ &2.33 &1.57 &0.99 &1.2(1) &\\
q &1.36 &0.5 &2.1 &-2(1) &\\
$\theta_0$, K &322 &1000 &1860 &1470(50)& \\ 
[0.5ex]
\hline

&Fe\textsuperscript{fcc} [5] &MgSiO$_{3}$\textsuperscript{ol} [6] &C\textsuperscript{gra} [7] &H$_{2}$O\textsuperscript{fluid} [4] &CH$_{4}$ [8]\\ 
$\rho_0$\textsuperscript{b} &8.17 &3.22(1)  &2.262(2) &0.998(5) &0.424 \\
$K_0$ &165.3 &130.0(9) &57.3(8) &3.0(5) & 7.85\\
$K'$  &5.5 &4.12(7) &4(fixed) &8.0(2) & 4(fixed) \\
[0.5ex]

\hline
&Fe$_7$C$_3$\textsuperscript{pm} [9] &MgSiO$_{3}$\textsuperscript{ppv} [10] &SiC\textsuperscript{3C} [11] &H$_{2}$O\textsuperscript{SI}  [4] & H$_{2}$ [12] \\ 
$\rho_0$\textsuperscript{b} &7.676(7) &4.03(1)   &3.214(1) &1.25(3) &7.86x10$^{-5}$ \\
$K_0$ &203(11) &219(5) &237(2) &8.0(6) &0.612 \\
$K'$  &8(1) &4(fixed) &4(fixed) &7.0(2) &6.813 \\
[0.5ex]

\hline
&$A$ &$B$ &$C$ &$D$ &$E$ \\
H$_{2}\textsuperscript{c}$ &17.633 & -6.33675 &0.0304574 &0.731393 &0.0085981 \\ 
\hline
    \end{tabular}%
 
{\raggedright
 \textsuperscript{a}All values are at ambient conditions (100 kPa and 298 K), except that the density of CH$_{4}$ is at 100 kPa and 30 K \citep{Mangan2017}. Uncertainties are on the last digits, if not listed.~~\textsuperscript{b}Assuming the densities of Fe\textsuperscript{bcc} \citep{komabayashi2010}, forsterite \citep{Finkelstein14Forsterite}, graphite \citep{Wang12Graphite}, and liquid H$_2$O \citep{Prakapenka21}, the stable phases at ambient conditions.~~\textsuperscript{c}Fitted polynomial \(V = AP^{-\frac{1}{3}} + BP^{-\frac{2}{3}} + CP^{-\frac{4}{3}} + (D+ET)P^{-1}\), where $P$ is pressure in GPa, and $T$ is temperature in K \citep{Tkacz2002}.~~Data sources: [1] hcp = hexagonal close packed  \cite{Smith2018_fe_eos}; [2] bdg = bridgmanite \cite{Fei2021}; [3] dia = diamond \cite{Bradley2009}; [4] Applies to ice VII, VII', and X \cite{Prakapenka21}; [5] fcc = face centered cubic \cite{komabayashi2010}; [6] ol = olivine (forsterite) \cite{Finkelstein14Forsterite}; [7] gra = graphite \cite{Wang12Graphite}; [8] \cite{Sun2009}. [9] pm = paramagnetic \cite{Chen12}; [10] ppv = post-perovskite \cite{Shieh06ppv}; [11] SI = super-ionic fluid \cite{Prakapenka21}; [12] \cite{Loubeyre1996}.}
\end{table}

As the fate of soot in planets and the relevant phases at applicable conditions are unknown, we estimated the density of soot at the reference condition of 1 bar and 300 K based on the linear correlation between the density and mean atomic number (Fig.~\ref{fig:rho_z}), without considering its thermodynamic stability. The thermoelastic properties of soot are bounded by the highly compressible water ice and the highly incompressible diamond (Fig.\ref{fig:eos}). These simplifying assumptions allow us to bound the M-R relations of soot planets, for comparisons with observed exoplanets with uncertainties of 10\% in radius and 20\% in mass.

\begin{figure}[ht]
\centering
    \begin{subfigure}{0.5\textwidth}
        \centering
        \includegraphics[width=0.95\linewidth]{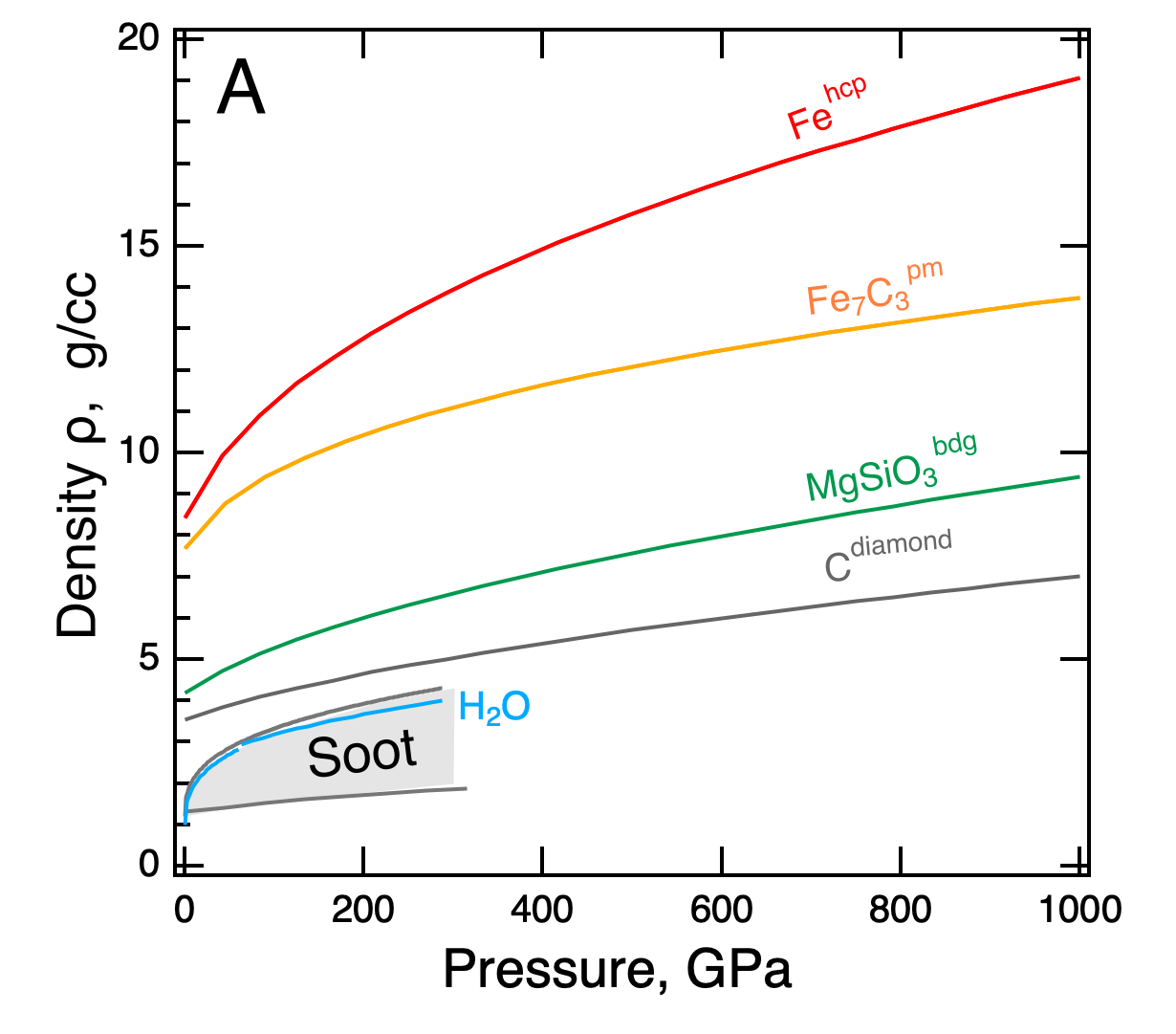}
    \end{subfigure}
    \hskip -0.3in
%\hfill
     \begin{subfigure}{0.5\textwidth}
         \centering
         \includegraphics[width=0.95\linewidth]{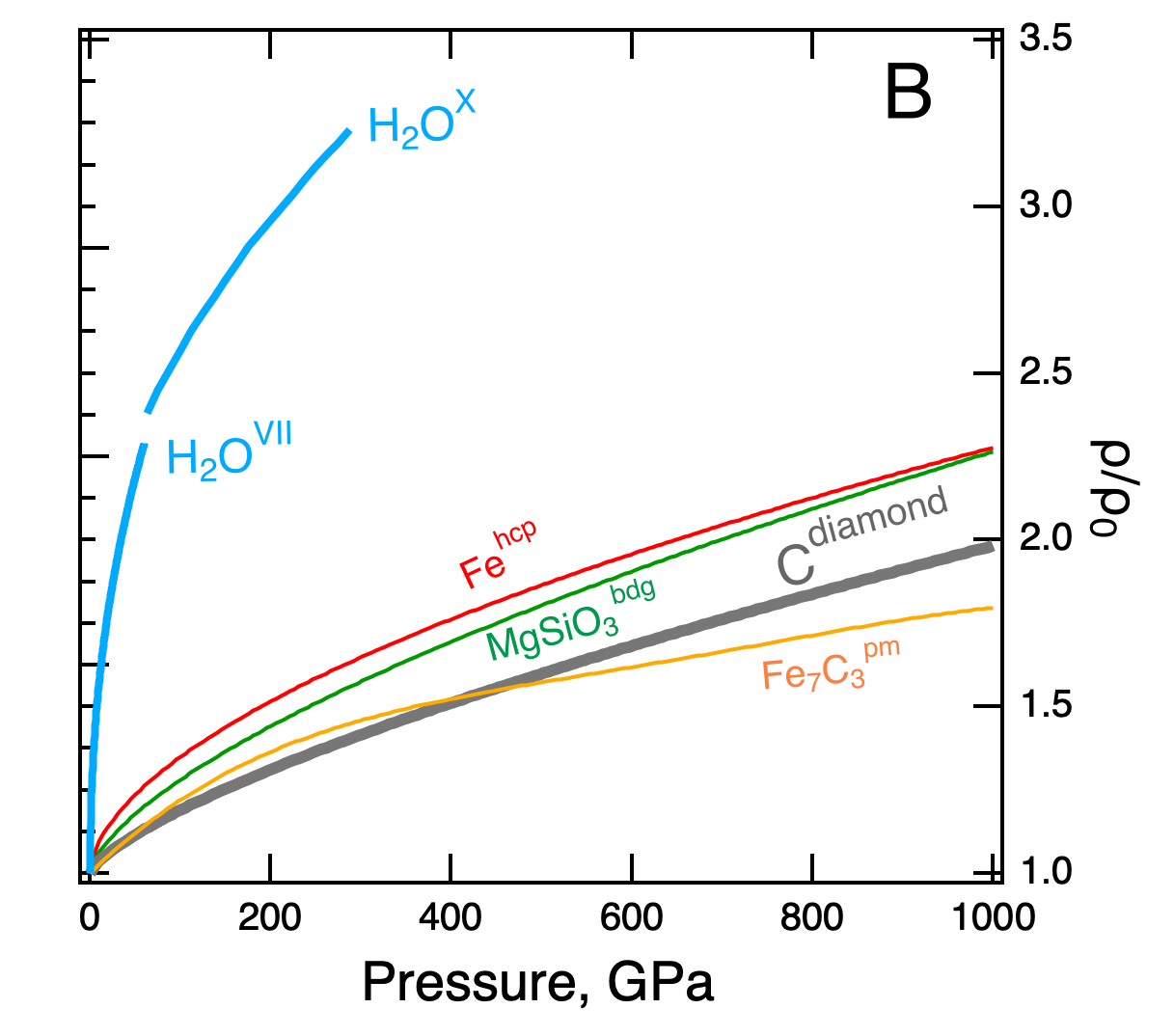}
    \end{subfigure}
\caption {(A) Compression curves of selected phases at 300 K. The compression curves of iron (Fe\textsuperscript{hcp}, red), iron carbide (Fe$_7$C$_3$\textsuperscript{pm}, orange), silicate (MgSiO\textsuperscript{bdg}, green), diamond (thin dark gray), and H$_2$O ice\textsuperscript{VII} (blue) are calculated using the Rose-Vinet equations-of state (Table~\ref{Tab:eos}). The shaded region represents bounds on the compression curve of soot calculated with the compressibilities of either diamond or H$_2$O ice\textsuperscript{VII}.  (B) Compression ratios $\rho$/$\rho_0$ of the same phases as in (A). Thick curves for diamond and H$_2$O ice\textsuperscript{VII} highlight that they are used to bound the compressibility of soot. }
\label{fig:eos}
\end{figure}

Our calculations involve several simplifying approximations and assumptions. The effects of phase transitions and thermal expansion are ignored.  Furthermore, in our model planets, all Fe is assumed to exist as metal in the core, and silicate layers are Fe-free. In real planets, the silicate layers contain some fraction of iron oxides and the cores contain some fraction of  Si, O as well as light elements (e.g., C, H, S, N). To assess uncertainties in the calculated M-R relations, we now consider the thermal profile of exoplanets and compare the compression curves of iron, MgSiO$_3$ silicate, diamond, water ice, and hydrogen at 300 K and high temperatures (Fig.~\ref{Fig:eos_thermal}).

\begin{figure}[ht]
\centering
    \begin{subfigure}{0.3\textwidth}
        \centering
        \includegraphics[width=1.0\linewidth]{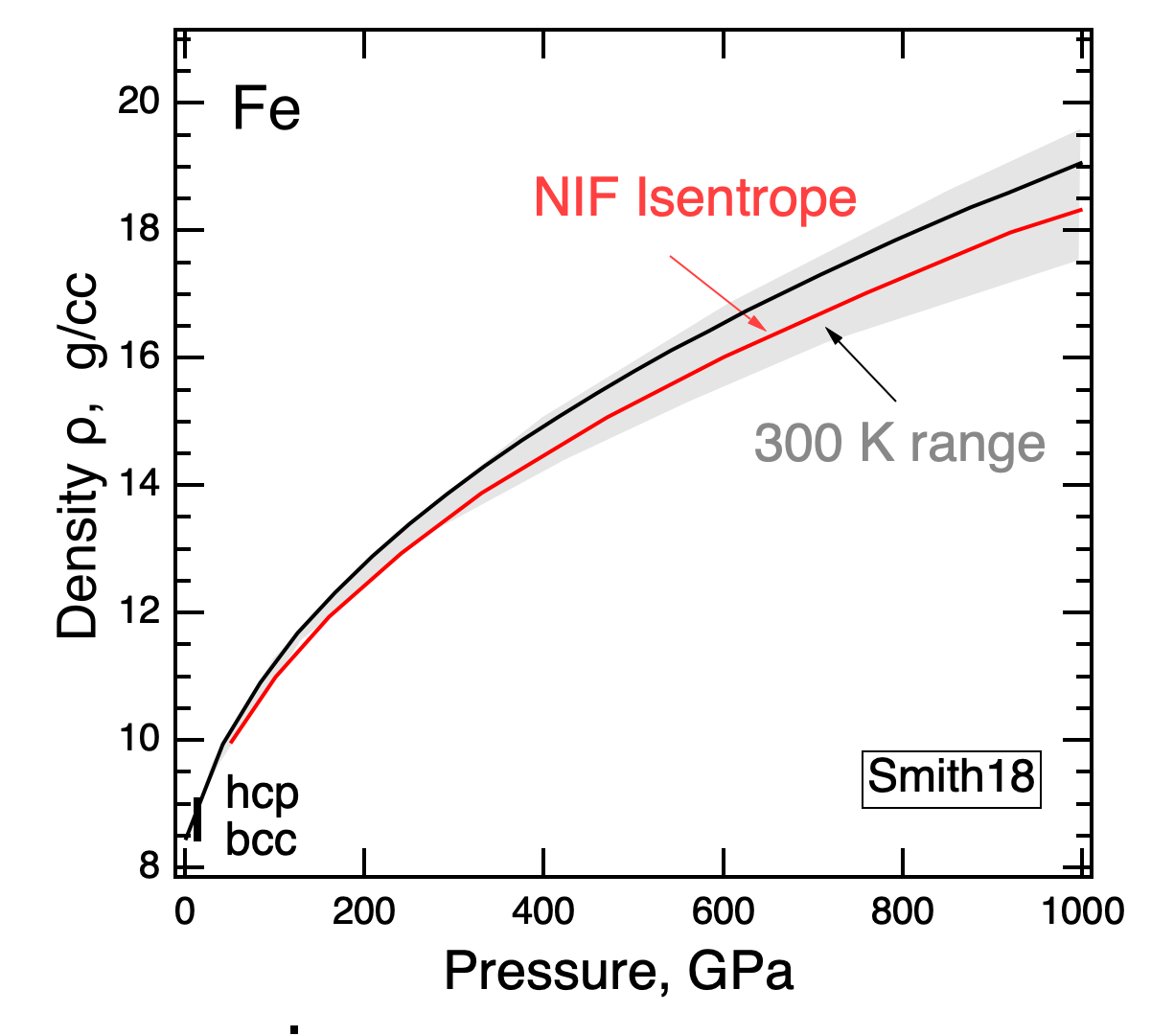}
    \end{subfigure}
    %\hskip 1.0in
%\hfill
     \begin{subfigure}{0.3\textwidth}
         \centering
         \includegraphics[width=1.0\linewidth]{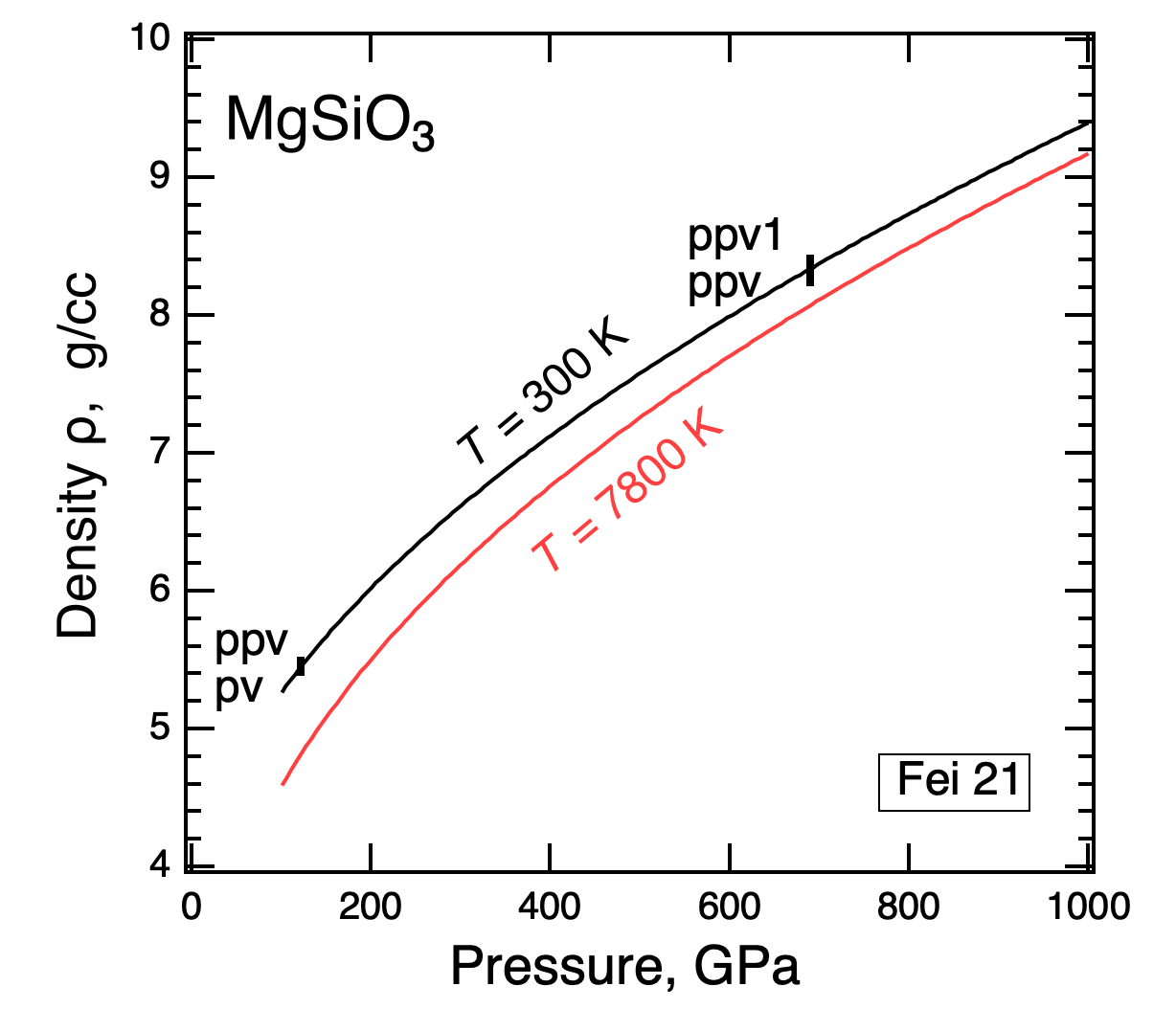}
    \end{subfigure}
%\hskip -0.3in
%\hfill
\begin{subfigure}{0.3\textwidth}
         \centering
         \includegraphics[width=1.0\linewidth]{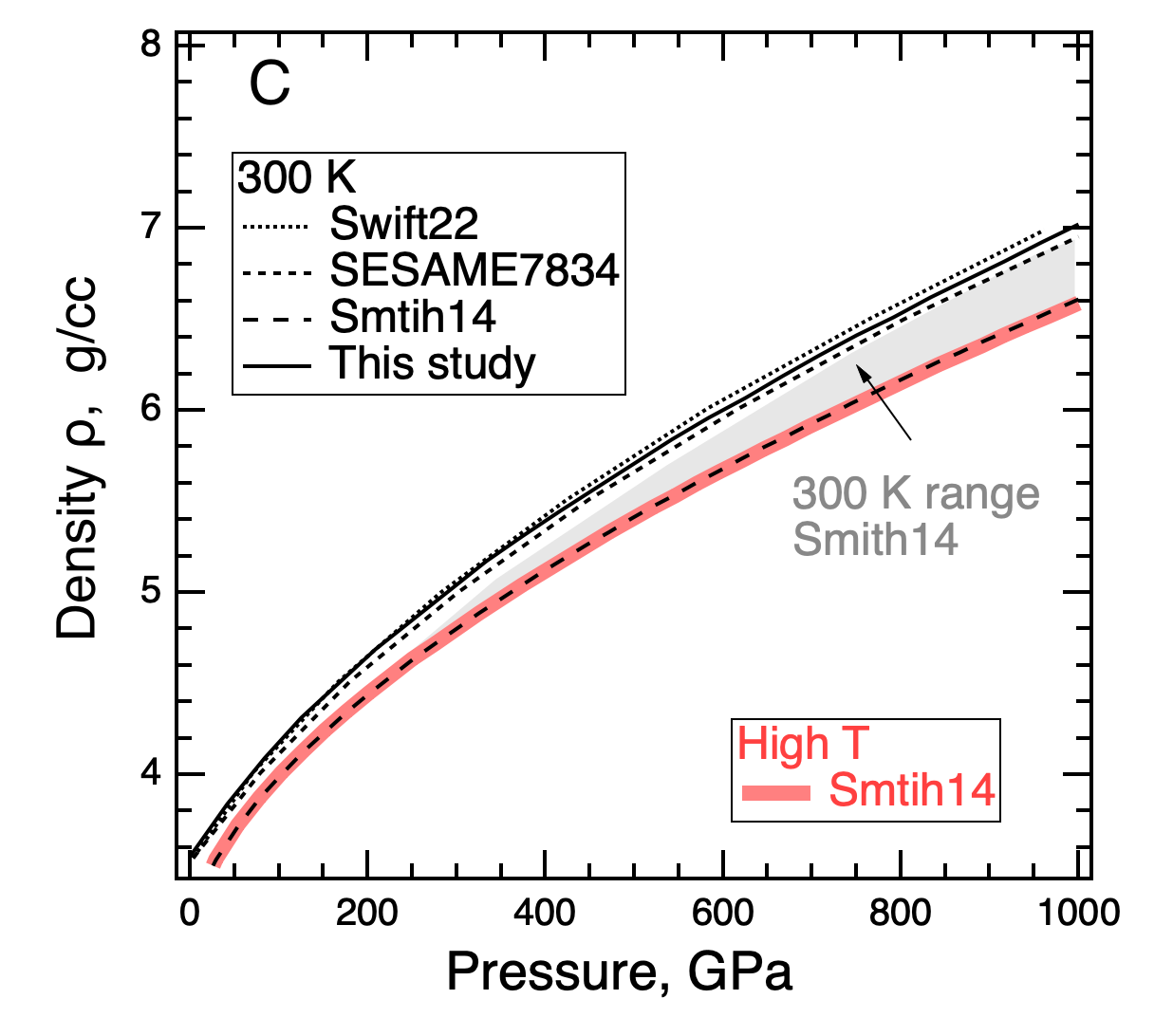}
    \end{subfigure}
    %\hskip -0.3in
\begin{subfigure}{0.3\textwidth}
         \centering
         \includegraphics[width=1.0\linewidth]{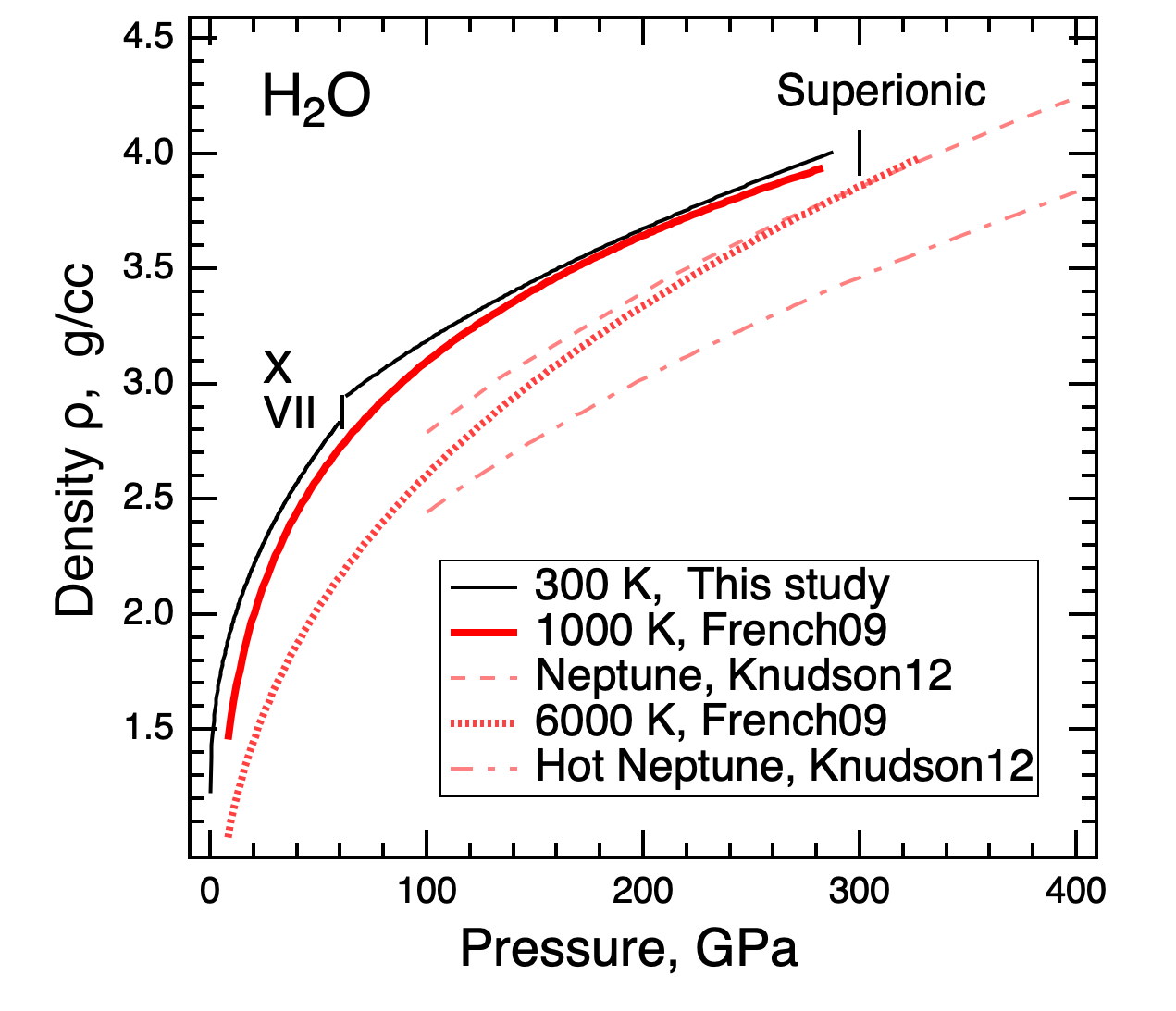}
    \end{subfigure}
    %\hskip -0.3in
%\hfill
     \begin{subfigure}{0.3\textwidth}
         \centering
         \includegraphics[width=1.0\linewidth]{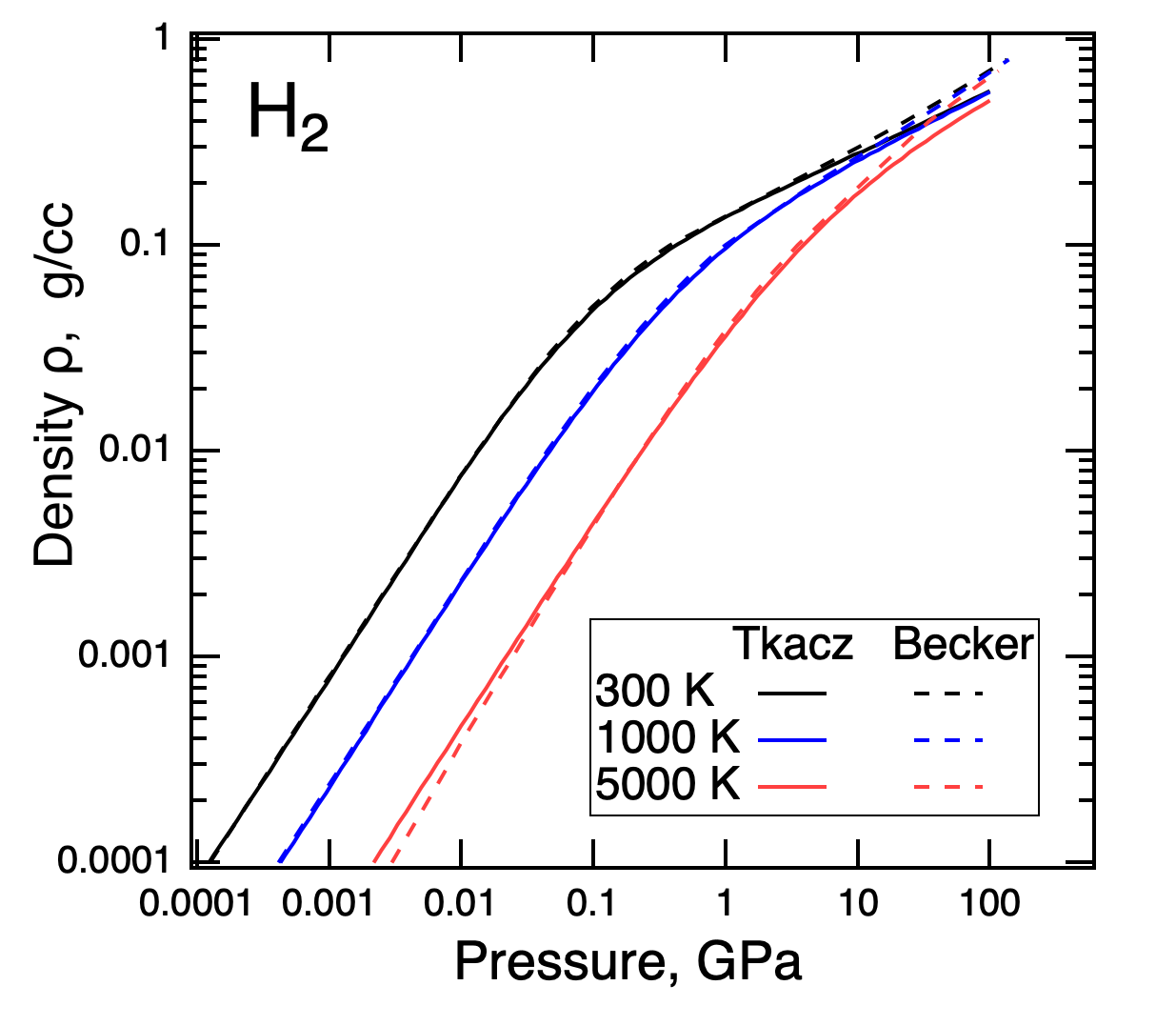}
    \end{subfigure}
    \caption{(A) The compression curves of iron at 300~K span a range that encloses the isentropic path of iron \protect\citep{Smith2018_fe_eos}. The vertical bar marks the density jump across the bcc--hcp transition at 14~GPa and 300~K.  (B) Comparison of isothermal compression curves of MgSiO$_3$ at 300~K and 7800~K \protect\citep{Fei2021}. Vertical bars mark the density jumps across the perovskite (pv) to post-perovskite (ppv) transition, and the ppv to ppv1 transition. (C) Comparison of 300~K compression curves \protect\citep{2022arXiv220316065S, Smith2018_fe_eos} and high-temperature compression curve \protect\citep{Smith2018_fe_eos} of diamond. In one study \protect\citep{Smith2018_fe_eos}, the 300~K range (gray) overlaps with the high-temperature (high-$T$) curve. (D) Compression curves of H$_2$O at 300~K (this study), compared with isothermal curves at 1000~K and 6000~K \protect\citep{2009PhRvB..79e4107F}, and with density--pressure profiles from the interiors of Neptune and hot Neptunes \protect\citep{2012PhRvL.108i1102K}. (E) Comparison of hydrogen equations of state (EoS). The empirical TEoS from metallurgical literature \protect\citep{Tkacz2002} can be used to calculate the mass--radius relation of planets with H$_2$ atmospheres. Although not designed for extreme pressures, the polynomial fit reproduces tabulated H-REOS.3 data \protect\citep{2014ApJS..215...21B} (based on \textit{ab initio} calculations for astrophysical use). Discrepancies between the two EoS sets are negligible above 20~GPa, but grow at low volumes (high pressures $>$100~GPa) due to the inverse relation between density and volume.}
\label{Fig:eos_thermal}
\end{figure}

\begin{comment}
The equilibrium temperature of an planet is governed by its distance to host star (e.g. \citep{Zeng2019}). The equilibrium temperature of a planet is given by:
\[
T_{\text{eq}} = T_{\star} \left( \frac{R_{\star}}{2D} \right)^{1/2} (1 - A_B)^{1/4}
\]
where $T_{eq}$ is the planet's equilibrium temperature, $T_{\star}$ is the stellar effective temperature, $R$ is the radius of the star, $D$ is the orbital distance between the star and planet, $A_B$ is the Bond albedo of the planet.
\end{comment}

The interiors of planets are expected to be hot due to adiabatic compression and partial conversion of the energy of accretion and differentiation into heat. Phase transitions and chemical reactions are common under the pressure and temperature conditions of planetary interiors (Fig.~\ref{Fig:eos_thermal}). For example, methane (CH$_4$) undergoes a series of phase transitions at high pressures and 300 K \citep{Sun2009}. Theory predicts that methane dissociates into ethane (C$_2$H$_6$) at 95 GPa, then butane (C$_4$H$_{10}$) at 158 GPa, and further, diamond (C) and hydrogen (H$_2$) above 287 GPa at 0 K \citep{Gao2010methane}. 

%The adiabatic temperature gradient in a convecting planetary interior can be expressed in terms of pressure as:
%\[
%\left( \frac{dT}{dP} \right)_{\text{ad}} = \frac{\gamma T}{\rho c_p}
%\]

The internal thermal profiles of exoplanets are highly uncertain, as they depend on the complex processes of accretion and early differentiation, the retention factors of formation energies, as well as the composition, long-term evolution, and age of planets (\citep{White25} and references therein). 

\begin{comment}
To evaluate the thermal effects on density, we can apply the Mie-Grüneisen-Debye model
\begin{equation}
    P(V, T) = P_{\text{cold}}(V) + P_{\text{thermal}}(V, T)
\end{equation}

where \( P_{\text{cold}}(V) \) is the cold pressure from the Vinet EOS, and the thermal pressure is:
\begin{equation}
    P_{\text{thermal}}(V, T) = \frac{\gamma(V)}{V} E_{\text{thermal}}(V, T)
\end{equation}

The thermal energy is:
\begin{equation}
    E_{\text{thermal}}(V, T) = 9nRT \left( \frac{T}{\Theta_D(V)} \right)^3 \int_0^{\Theta_D(V)/T} \frac{x^3}{e^x - 1} dx
\end{equation}

where $\gamma(V)$ is Grüneisen parameter, $\Theta_D(V)$is Debye temperature, $n$ is the number of atoms per formula unit, $R$ is gas constant, $T$ is  temperature in K, and $V$ is specific volume.

The volume dependence for the Grüneisen parameter is
\begin{equation}
    \gamma(V) = \gamma_0 \left( \frac{V}{V_0} \right)^q
\end{equation}

and likewise for the Debye temperature:

\begin{equation}
    \Theta_D(V) = 
    \begin{cases}
    \Theta_{D0} \exp\left[ -\frac{\gamma_0}{q} \left( \left( \frac{V}{V_0} \right)^q - 1 \right) \right], & q \ne 0 \\
    \Theta_{D0} \left( \frac{V}{V_0} \right)^{-\gamma_0}, & q = 0
    \end{cases}
\end{equation}
\end{comment}

There are limited constraints on material properties under relevant conditions (Table~\ref{Tab:eos}). Existing data suggest that the influences of phase changes and thermal expansion on densities are often smaller than the uncertainties in the 300-K compression curves at high pressures(Fig.~\ref{Fig:eos_thermal}). Fortunately, thermal expansion decreases rapidly with pressure, and therefore the expected effects are limited both at low pressures where temperatures are relatively low, and at high pressures where the thermal expansion coefficient is small. Phase transitions could lead to densification or density reduction. While pressure-induced transitions into denser structures partially offset that of thermal expansion on densities, heating would induce transitions into phases of lower densities. In addition, material structure at very high pressure is mainly supported by the electron degeneracy pressure, which has similar functional dependence on compression for different materials, and therefore the high-pressure density chiefly depends on the mean atomic number and secondarily on the crystal structure or temperature \citep{Zeng2019}.

%And the adiabatic gradient in radius is:
%\[
%\frac{dT}{dr} = -\frac{g \gamma T}{c_p}
%\]
%where:
%\begin{itemize}
%  \item \( T \) is the temperature,
%  \item \( P \) is the pressure,
%  \item \( r \): radius
%  \item \( \gamma \) is the Grüneisen parameter,
%  \item \( \rho \) is the density,
%  \item \( c_p \) is the specific heat at constant pressure.
%\end{itemize}

%\bigskip

%\newpage
%\clearpage
%\vspace{\baselineskip}. 

Previous studies considered phase and temperature changes for low-mass planet models \citep[e.g.,][]{Valencia2007ApJ...665.1413V, Sotin2007Icar..191..337S}, Although more realistic, they involve complicated interior boundary conditions and highly uncertain EoS, and yielded M-R relations that agree with our simplified models (Fig.~\ref{Fig:m_r_thermal}).

\begin{figure}[ht]
    \centering
  \includegraphics[width=0.5\linewidth]{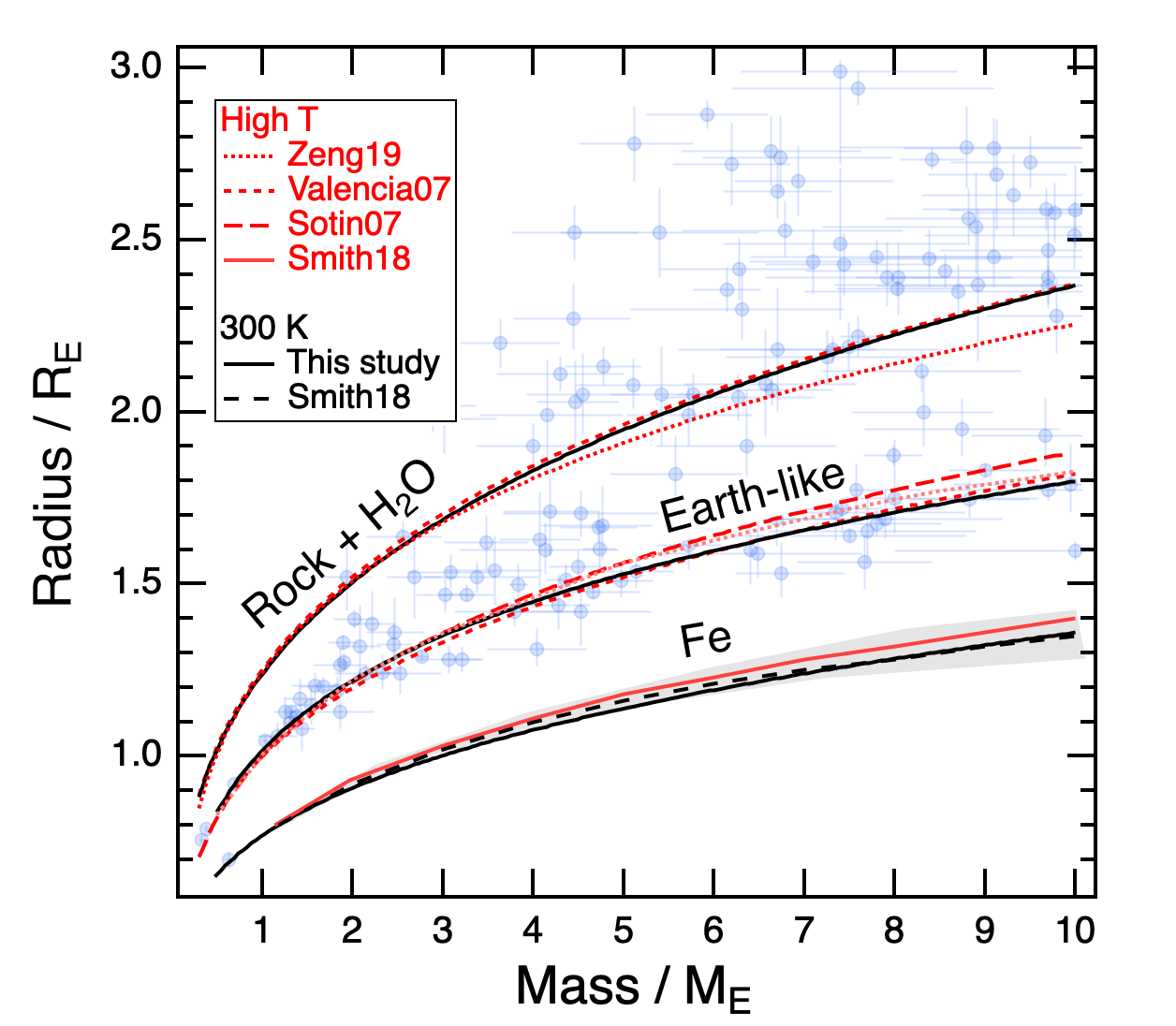}
   \caption{Comparison of mass--radius relations for iron planets, Earth-like rocky planets, and water-worlds composed of 50\% rock and 50\% water. Curves from models that include phase transitions and thermal effects are shown in red (dotted: \protect\citep[][]{Zeng2019}; short-dashed: \protect\citep[][]{Valencia2007ApJ...665.1413V}; long-dashed: \protect\citep[][]{Sotin2007Icar..191..337S}; solid: \protect\citep[][]{Smith2018_fe_eos}). Models computed at 300~K are shown in black (solid: this study; dashed: \protect\citep[][]{Smith2018_fe_eos}).}
    \label{Fig:m_r_thermal}
\end{figure}

%\newpage

%\begin{figure}[ht]
%\centering
%\includegraphics[width=\linewidth]{R_M_PurePlanet.png}
%\caption{ Mass-radius relations of model planets made of a single component. The curves of Fe (red), MgSiO$_3$ (green), C (thin dark gray), H$_2$O (blue), and CH$_4$ (thin ligh5 gray) are calculated using the equation-of-state parameters in Table S3. The gray shaded region represent the range for pure soot planet bounded by two end-member scenarios of highly incompressible diamond and highly compressible water ice. The soot curves (thick gray) are calculated using the density of soot at 100 kPa (1 bar) and 300 K (Table S2) and the compressibilities of diamond and H$_2$O\textsuperscript{VII} (Table S3). The observed mass and radius of selected exoplanets (blue dots with error bars, \citep{Southworth2011}) are plotted for comparison.}

%(explain radius gap histogram, with reference).
  
%\label{fig:S3}
%\end{figure}

%\newpage
%\bibliographystylesupp{plain}
%\bibliographysupp{z}

\end{document}